\documentclass{emulateapj}
\usepackage{amsmath}

\shorttitle{Binary black hole  mergers from field triples}
\shortauthors{Antonini et al.}

\begin{document}
\def\gap{\;\rlap{\lower 2.5pt
\hbox{$\sim$}}\raise 1.5pt\hbox{$>$}\;}
\def\lap{\;\rlap{\lower 2.5pt
 \hbox{$\sim$}}\raise 1.5pt\hbox{$<$}\;}

\newcommand\MBH{\rm BH}
\newcommand\NC{NSC}
\title{Binary black hole mergers from field triples: properties, rates and the impact of stellar evolution}

\author{Fabio~Antonini$^{1,*}$, Silvia Toonen$^{2,*}$, Adrian S. Hamers$^{3}$}
\affil{*Contributed equally to this work. \\
(1) Center for Interdisciplinary Exploration and Research in Astrophysics (CIERA)
and department of physics and astrophysics, 
Northwestern University; (2) Astronomical Institute Anton Pannekoek, University of Amsterdam, P.O. Box 94249, 1090 GE, Amsterdam; (3) Institute for Advanced Study, School of Natural Sciences, Einstein Drive, Princeton, NJ 08540, USA.}

\begin{abstract}
We consider the formation of binary black hole mergers through the evolution of field massive triple stars. 
In this scenario, favorable conditions for the inspiral of a black hole binary are 
initiated  by its gravitational  interaction with 
a distant companion, rather than by 
a common-envelope phase  invoked in standard binary evolution models.
We use a code that follows self-consistently the evolution of massive triple stars, combining 
 the secular triple dynamics (Lidov-Kozai cycles) 
with stellar evolution.
 After a black hole triple is formed, 
its  dynamical evolution
 is  computed using either the orbit-averaged equations of motion, 
 or a high-precision direct integrator for triples with  weaker hierarchies for which 
 the secular perturbation theory breaks down.
Most black hole mergers in our models are 
 produced in the latter non-secular dynamical regime.
 We derive the properties of the merging binaries and compute a black hole merger rate in 
 the range $(0.3- 1.3)\ \rm Gpc^{-3}yr^{-1}$, or up to
$\approx 2.5\ \rm Gpc^{-3}yr^{-1}$ if the black hole orbital planes
  have initially random orientation. 
Finally, we show that black hole mergers
 from the triple channel have significantly higher eccentricities than those 
 formed through the evolution of massive binaries or in dense star clusters.
Measured eccentricities could therefore
be used to uniquely identify binary mergers formed through the evolution of  triple stars.
While our results suggest up to $\approx10$  detections per year with Advanced-LIGO, 
the high eccentricities  could render  the merging binaries harder to
detect  with planned space based interferometers such as LISA.
   \end{abstract}

\section{ Introduction}
The recent breakthrough detection of gravitational waves (GWs) from merging black hole (BH) binaries by Advanced-LIGO (aLIGO)
has generated enormous interest in understanding how these sources form \citep{2016PhRvL.116f1102A,2016ApJ...818L..22A,2016PhRvL.116x1103A}.  
With the many hundreds to thousands
of merging binary signals now expected to be detected over the next years,
the numerical and analytical modeling  of
the formation of compact object binaries will be central for a correct astrophysical interpretation of the 
GW sources we are about to discover \citep{2016arXiv160203842A}.

Several channels for the formation of BH binary mergers have been proposed.
 Binary BHs can form as the result of: (i) the evolution of isolated massive
  binaries in galactic fields \citep{2010ApJ...715L.138B,2016ApJ...819..108B,2016MNRAS.458.2634M};
 (ii) dynamical interactions in galactic nuclei, with and without a massive BH \citep{2012ApJ...757...27A,2016ApJ...831..187A};
(iii) dynamical exchange interactions in the dense stellar core of globular clusters \citep{2015PhRvL.115e1101R,2016ApJ...832..192H,2017ApJ...836L..26C}
or young massive star clusters \citep{2017MNRAS.tmp...20B};
(iv) the evolution of isolated triples in galactic fields \citep{2016arXiv160807642S}, leading  to a merger of the inner binary through the
 Lidov-Kozai
mechanism \citep[LK;][]{1962P&SS....9..719L,1962AJ.....67..591K} coupled 
with energy loss due to GW radiation \citep[e.g.,][]{2011ApJ...741...82T,2014ApJ...781...45A,2016MNRAS.463.2443K}.

In this paper we make precise predictions for scenario (iv).
 We use a numerical algorithm \citep{2016ComAC...3....6T}
that combines the secular triple dynamics with stellar evolution and interactions to
follow the evolution of stellar triples and the way this results in the
 formation of BH triples. After their formation, the BH triples
 are evolved forward in time using either the octupole level secular equations of motion of triple systems 
  \citep[e.g.,][]{2002ApJ...578..775B,2013MNRAS.431.2155N} or a high-precision direct integrator for systems 
with   weaker hierarchies where 
 the secular equations become inaccurate \citep{2008AJ....135.2398M}.
  Unlike previous studies \citep[e.g.,][]{2016arXiv160807642S},
  our models  follow the  stellar, binary and triple evolution  self-consistently together 
 with the triple secular dynamics,
prior to the formation of the triple BH system. 
 We perform a population synthesis study  that allows us 
to make concrete predictions about the merger rate and properties of BH binaries formed in massive field triples.

The paper is organized as follows. In Section \ref{mtd}, we describe the numerical method and 
in Section \ref{ICs}, we describe our choice for the initial conditions. In Section \ref{results}, we present  the main results 
of our simulations and compute 
the properties and merger  rate of binary BHs formed in our models.
In Section\ \ref{ac} we discuss the importance of the non-secular dynamical regime and possible effects related to encounters with field stars.
 Section \ref{concl} sums up.

 \section{ Method}\label{mtd}
 To simulate the formation of triple BH systems, we use the recently published code \texttt{TRES} \citep{2016ComAC...3....6T}
  for studying the evolution   of coeval, dynamically stable, hierarchical, stellar triples. 
The simulations start with three stars on the zero-age main sequence in a specific orbital configuration. Consequently the 
evolution of the stars and the orbit is followed in time. Stellar evolution is modeled in a parametrized way through the fast stellar 
evolution code SeBa \citep{2012A&A...546A..70T}. 
The parametrization is originally based on analytic formulae that are fitted to detailed stellar evolution 
calculations of single stars \citep{2000MNRAS.315..543H}. 
Mass loss from stellar winds is incorporated as in \citet[][]{2013A&A...557A..87T}. 
The wind mass loss rates are time-dependent and reflect the stellar phase and parameters (e.g. luminosity and radius). 

 The secular orbital evolution of the triple is modeled by solving a set of ordinary differential equations (ODEs) for the inner and outer orbital 
 elements and the spin frequencies of the bodies, similarly to 
\citet{2013MNRAS.430.2262H}. In addition to the secular three-body dynamics, we include 
tidal evolution (dissipation and precession due to tidal bulges) in the inner binary assuming the equilibrium tide model and stellar spins parallel to the orbit 
\citep{1981A&A....99..126H}, general relativistic corrections (precession at the 1PN order, and gravitational wave emission at the 2.5PN order), and orbital changes associated with stellar winds. During the ODE integration we check for dynamical instability using the \citet{2001MNRAS.321..398M} stability criterion, and for the onset of mass transfer using 
the results of \citet{2007ApJ...667.1170S}  (see also \citet{2016ApJ...825...70D,2016ApJ...825...71D}). The latter can be triggered by eccentricity 
driving by LK oscillations, although the subsequent evolution is not taken into account here; instead, 
we track the onset of mass transfer in our simulations and do not model the further evolution.
 
 \citet{2014ApJ...781...45A} show that the angular momentum of 
the inner binary can change by order of itself in one orbital period if:
\begin{equation}\label{aa}
\sqrt{1-e_{\rm 1}} \lesssim \sqrt{1-e_{\rm crit}}\equiv 5\pi {m_3\over m_{\rm 1}+m_{\rm 2}} 
\left[ a_1 \over a_2(1-e_2)\right]^3~.
\end{equation}
where $e_1$ ($e_2$), and $a_1$ ($a_2$) are the eccentricity and semi-major axis of the inner (outer) binary;
$m_1+m_2$ and  $m_3$ are the total mass of the inner binary 
and the mass of the tertiary companion, respectively.
In what follows we refer to the high-$e$ region of parameter space defined by Equation\ (\ref{aa}) 
as the ``non-secular'' dynamical regime.
 If the eccentricity of the inner binary becomes larger than 
$e_{\rm crit}$,  the
outer perturber can significantly change
the angular momentum of the eccentric binary at the last
apoapsis passage leading to a jump in angular momentum of order unity 
\cite[see Antonini \& Perets, and 
Eq. (7) in][]{2012arXiv1211.4584K}.
 In this situation the secular equations of motion \citep[e.g.,][]{2016ARA&A..54..441N} become a poor approximation.
For this reason,  during the secular integrations  we checked 
whether  $e_1$ became larger than $e_{\rm crit}$. If this happened,
we reintegrated the evolution of the BH triple
system using the  direct code AR-CHAIN  \citep{2008AJ....135.2398M}.
The AR-CHAIN code combines the use of the chain regularization
method and the time-transformed
leapfrog scheme to avoid singularities, and includes relativistic effects to the motion that
 are added as
 1, 2  and 2.5 order post-Newtonian corrections. 
We assumed that the orbital phases were initially random and ran
our direct integrations up to a maximum time of
$10{\rm Gyr}$. For most triples, this maximum time 
corresponds to  a few thousand times the 
LK timescale   \citep{1997Natur.386..254H}:
\begin{equation}\label{tkl}
T_{\rm LK}=P_1\left(m_{\rm 1}+m_2\over m_3\right) \left(a_2\over a_1\right)^3
\left(1-e_2^2 \right)^{3/2}\ ,
 \end{equation}
with $P_1$ the orbital period of the inner binary.

 \section{initial conditions}\label{ICs}
 
 \subsection{Orbit and mass distributions}
  The stellar triples in our simulations are initialized as described in what follows.
  In total, we consider 6 different models (see Table 1).

  In all our models, 
   we assume Solar metallicity and
 sample the mass of the most massive star in the inner binary from  
  a Kroupa  initial mass function \citep{2002Sci...295...82K}.
    We adopt  a flat mass ratio distribution between 
  $0$ and $1$
  for both the inner binary, i.e. $m_2/m_1$, and the outer binary, i.e. $m_3/(m_1+m_2)$. 
  This choice is consistent with  observations of massive binary stars
  which indicate a nearly flat distribution of the mass ratio  
  \citep[e.g.,][]{2012Sci...337..444S,2013ARA&A..51..269D,2014ApJS..213...34K}.
  Stellar masses were sampled in the range $[22 \rm M_{\odot},100\rm M_{\odot}]$ for $m_1$
and $m_2$, and in the range $[m_{3,\rm min},100\rm M_{\odot}]$ for the outer star.
   In models A1, B1 and C1 we set $m_{3,\rm min}= 0.1M_{\odot}$;
  in models A2, B2 and C2 we set  $m_{3,\rm min}= 22M_{\odot}$.  
  The upper limit on the mass comes from the fact that the stellar evolution tracks used in TRES are not valid above 
  $100M_\odot$.
   
 The distribution of orbital separations/periods  is often assumed to be flat in log-space  (\"{O}pik's law). 
 However,  \citet{2012Sci...337..444S} 
 find that the orbital periods follow a power-law function:
  $\log(P/{\rm days})^{-0.55}$, in the range $(0.15-5.5)$; while
for wide orbits, there are indications that the distribution is more similar to the canonical \"{O}pik's law. 
Hence, we take $N\propto \log(P_1/{\rm days})^{-0.55}$ for the period of the inner binary, 
and flat in log-space for the period of the outer orbit.

Any common envelope phase of evolution will likely 
  shrink the inner orbit so that the Lidov-Kozai mechanism will become more strongly
suppressed by the relativistic precession of the inner binary orbit \citep[e.g.,][]{2002ApJ...578..775B}.
For this reason, in order to make sure that the periapsis  distance  of  the
inner  binary is large enough that no common envelope phase or mass transfer occurs 
we impose  a  minimum orbital separation 
$a_1(1-e_1^2)\geq 2500R_\odot\approx 11\rm AU$.  
This latter is the maximum radius of a low-mass BH progenitor \citep[e.g.,][]{2013A&A...557A..87T}.
We have placed a conservative limit by using the semi-latus rectum instead of the periapsis of the orbit, 
because in this way we exclude binaries that would experience mass transfer even if they would be isolated, 
as tides would circularize the binary to the semi-latus rectum.   
We also imposed a maximum  separation, $a_{2,\rm max}$, for the outer star.
 In models A1, B1 and C1 we set  $a_{2,\rm max}= 5\times 10^6 R_\odot= 2.2\times10^4\rm AU$;
in models A2, B2 and C2 we set  $a_{2,\rm max}= 5\times 10^5 R_\odot= 2.2\times10^3\rm AU$.

For  the orbital eccentricities of the inner binary, $e_1$, and outer binary, $e_2$,  we adopt a flat distribution between 0 and 1.

Lastly, we sample the initial orbital inclination angle 
 between the inner and outer orbit, $I$, randomly 
in $\cos I$ with $-1<\cos I<1$,
 and the initial arguments  of periapsis of inner and outer orbits $g_1$ and $g_2$
 randomly between $0$ and $2\pi$.
 From the triple systems obtained,  we subsequently reject 
 those that are dynamically unstable based on the stability criterion of
\citet{2001MNRAS.321..398M}.

\begin{figure*}
\begin{center}
 \includegraphics[angle=0,width=6.8in]{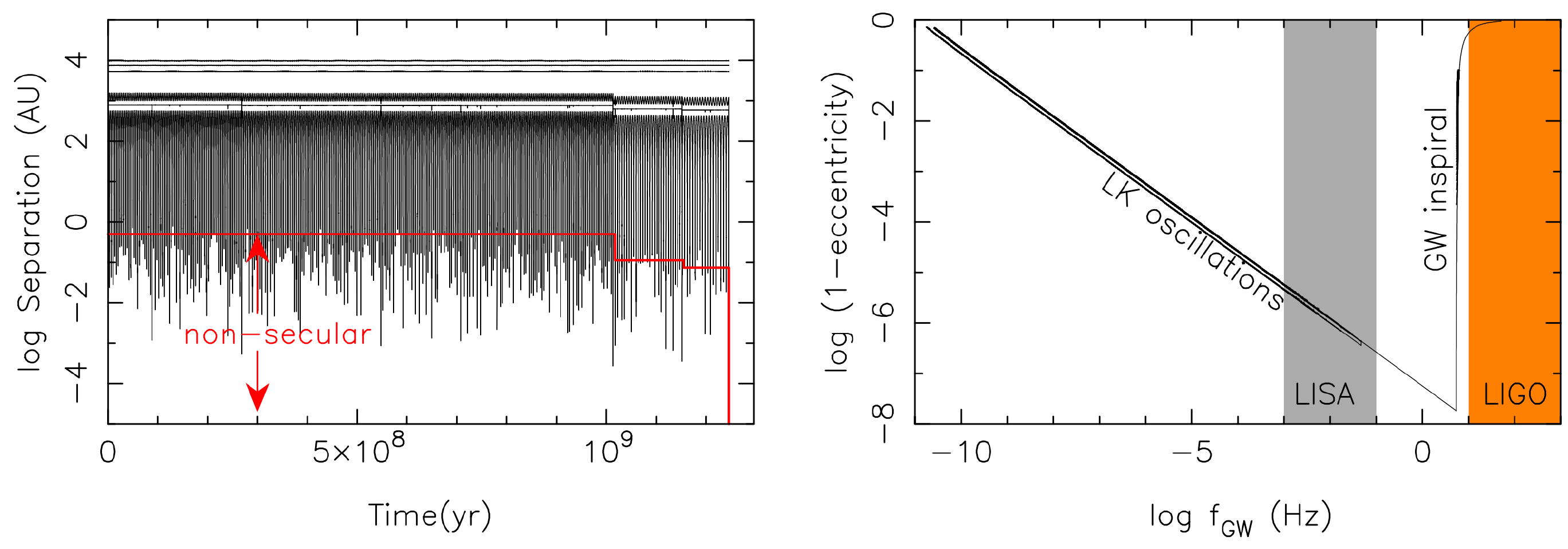} 
    \caption{ Formation of a BH merger  in one of our simulations 
    (AR-CHAIN integration).  Masses and initial orbital parameters were as follows:
    $m_1=8.96~M_{\odot}$,  $m_2=7.51~M_{\odot}$, $m_3=8.35~M_{\odot}$, 
    $a_1=1727\rm AU$, $a_2=16571\rm AU$, $e_1=0.65$, $e_2=0.29$,
    $I=93^{\circ}.8$, $g_1(\rm rad)=0.61$, $g_2(\rm rad)=-2.82$, and the longitude of the ascending node $\Omega_1(\rm rad)=-2.4$.
             Radial excursions of the three
BHs (semi-major axis, periapse and apoapse) are shown as functions of time, with the 
red line demarcating the region below which the standard secular perturbation theory breaks 
down according to Equation~(\ref{aa}). At $\approx 10^9\rm yr$, the two BHs
approach closely each other and energy dissipation due to GW radiation leads to 
a sudden decrease in the inner binary semi-major axis; at the end of the simulation, after 
$\approx 1.3\times 10^9\rm yr$, the inner binary merges.
The right panel displays the evolution 
of the inner binary eccentricity
as a function of the peak gravitational wave frequency as defined in Equation\ (\ref{fgw}).
The binary evolves into the LISA frequency band, $f_{\rm GW}\gtrsim 10^{-3}\rm Hz$ \citep{2017arXiv170200786A},
with extremely high eccentricities. The GW driven inspiral starts at $f_{\rm GW}\approx 5\rm Hz$, outside the LISA frequency window.
 By the time the binary  reaches 
the $10\rm Hz$ frequency its eccentricity is  $e_1\approx 0.4$.}
\label{example}
\end{center}
\end{figure*}

 \subsection{Kick distributions}
  Any asymmetry  in  the  supernova, such  as  in  the  mass  or  neutrino  loss  can  give  rise
to  a  natal-kick to  the  newly formed compact-object.
If the collapsing star is part of a binary or triple, the natal kick  alters the orbit.
If the overall kinetic  energy imparted through
the natal kicks is higher than the binding energy of the binary
orbit the supernova kick can fully unbind the system. 
We assume that the supernova is instantaneous and that the supernova-shell does not impact 
the companion star(s).

In our simulations the orbital evolution of a 
triple due to  an instantaneous supernova kick was modeled using the procedure described in 
 the appendix of \citet{2016ComAC...3....6T}.
 In summary, the positions and velocities of the three bodies before and just after the supernova are computed from the orbital elements assuming a random 
mean anomaly, and assuming no changes in the position vectors. To the body undergoing the supernova, the natal kick velocity is added to the orbital velocity and the mass is updated. Consequently, we compute the new eccentricity and specific angular momentum vectors for the inner and outer orbits, and the new orbital elements are computed from these vectors. Note that the total orbital angular momentum vector of the triple is {\it not} conserved because of the mass loss and the natal kick velocity. Therefore, the reference plane for the orbital elements (i.e., the invariable plane perpendicular to the total orbital angular momentum vector) changes due to the supernova, and our approach takes this effect into account.

The distribution of natal kick velocities of BHs  is unknown.
 We  therefore consider  a number of choices for the natal kick velocity (see Table 1).
 In models A1 and A2, we  assume no natal kick during black hole formation.
   We note that even in this case, the binary receives a kick to its center of mass because one of the
massive components suddenly loses mass \citep{1961BAN....15..265B}. 

In models B1, B2, C1 and C2, we consider a non-zero kick velocity for the newly formed BHs.
We implement momentum-conserving kicks,
in which we assume that the momentum imparted on a BH is the same
as the momentum given to a neutron star.  Thus,
the kick velocities for the BHs will be reduced with respect to those of neutron stars
by a factor   $m_\mathrm{NS}/m_{\rm BH}$, with $m_\mathrm{NS}=1.4M_{\odot}$ the typical neutron 
star mass, and $m_{\rm BH}$ the mass of the forming BH.
The neutron star kick distributions 
we adopt  are constrained by observations.
In models B1 and B2, we use the neutron star kick distribution of \citet{2005MNRAS.360..974H}, which
can be approximated by a Maxwellian distribution with dispersion $265\rm km\ s^{-1}$ for neutron stars,
and $ \approx 40 \rm km\ s^{-1}$ for BHs  after 
the kick velocities have been reduced by the ratio of BH to neutron star mass.
  In models C1 and C2, we use the neutron star kick distribution of \citet{2002ApJ...568..289A}.
 This latter distribution is bimodal with   a peak  at $\approx 100\rm km\ s^{-1}$ and a lower peak 
 at $\approx  700\rm km\ s^{-1}$ for neutron stars; this converts  into a   kick distribution for BHs 
 with a main peak at $\approx 20\rm km\ s^{-1}$ and a lower peak 
 at $\approx 120\rm km\ s^{-1}$.

 Finally, in all our models, we assume that stars with a zero-age main-sequence mass 
     $m\geq 40\rm M_{\odot}$ do not receive any  natal kick.
  This is consistent with the direct
collapse scenario described in \citet{2001ApJ...554..548F},
where the
fallback completely damps any natal kick for $m\gtrsim 40\rm M_{\odot}$, assuming Solar metallicity.

\begin{figure*}
\begin{center}
\raisebox{0.mm}{ \includegraphics[angle=270,width=2.4in]{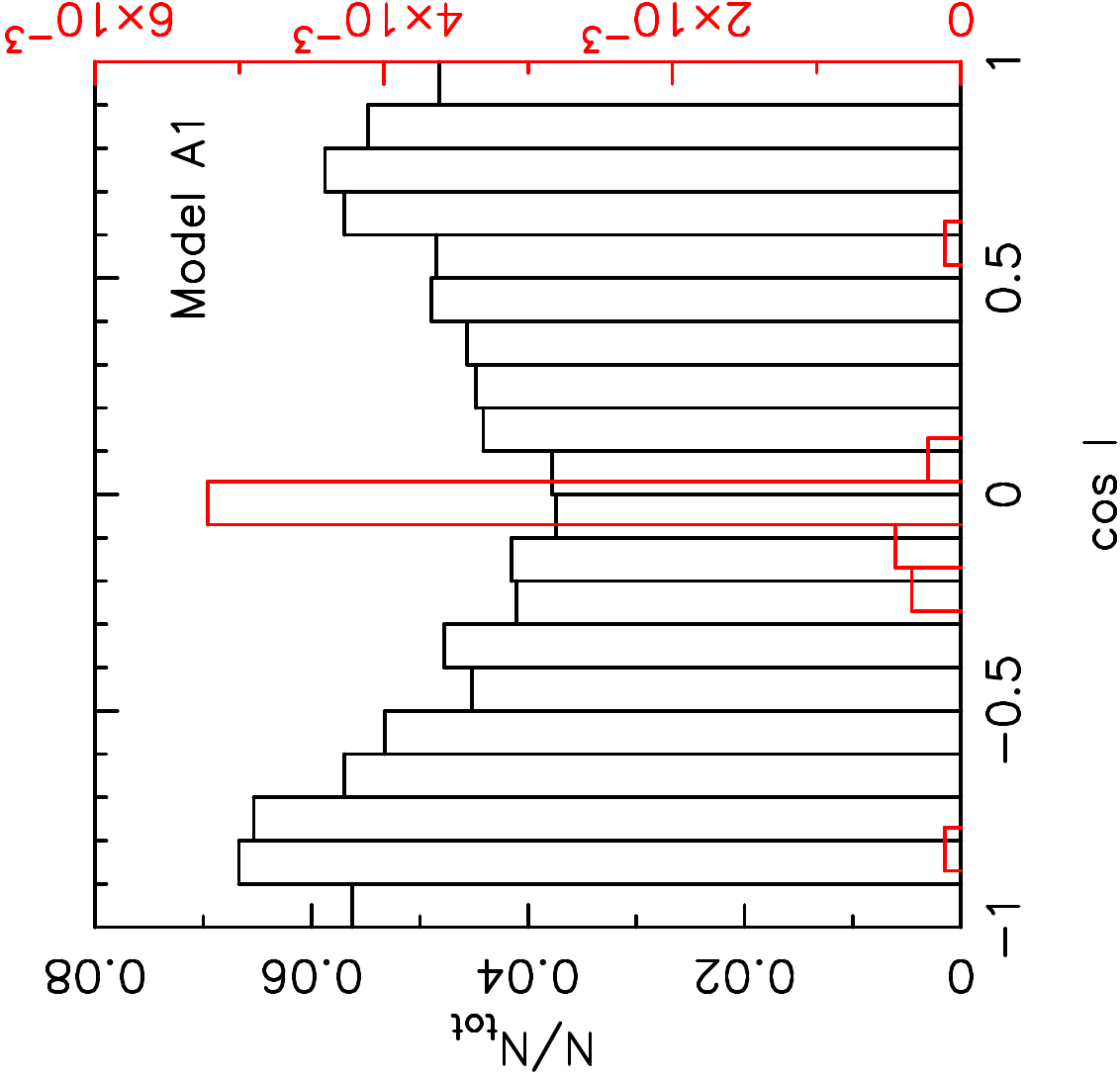}  }
 \includegraphics[angle=270,width=2.4in]{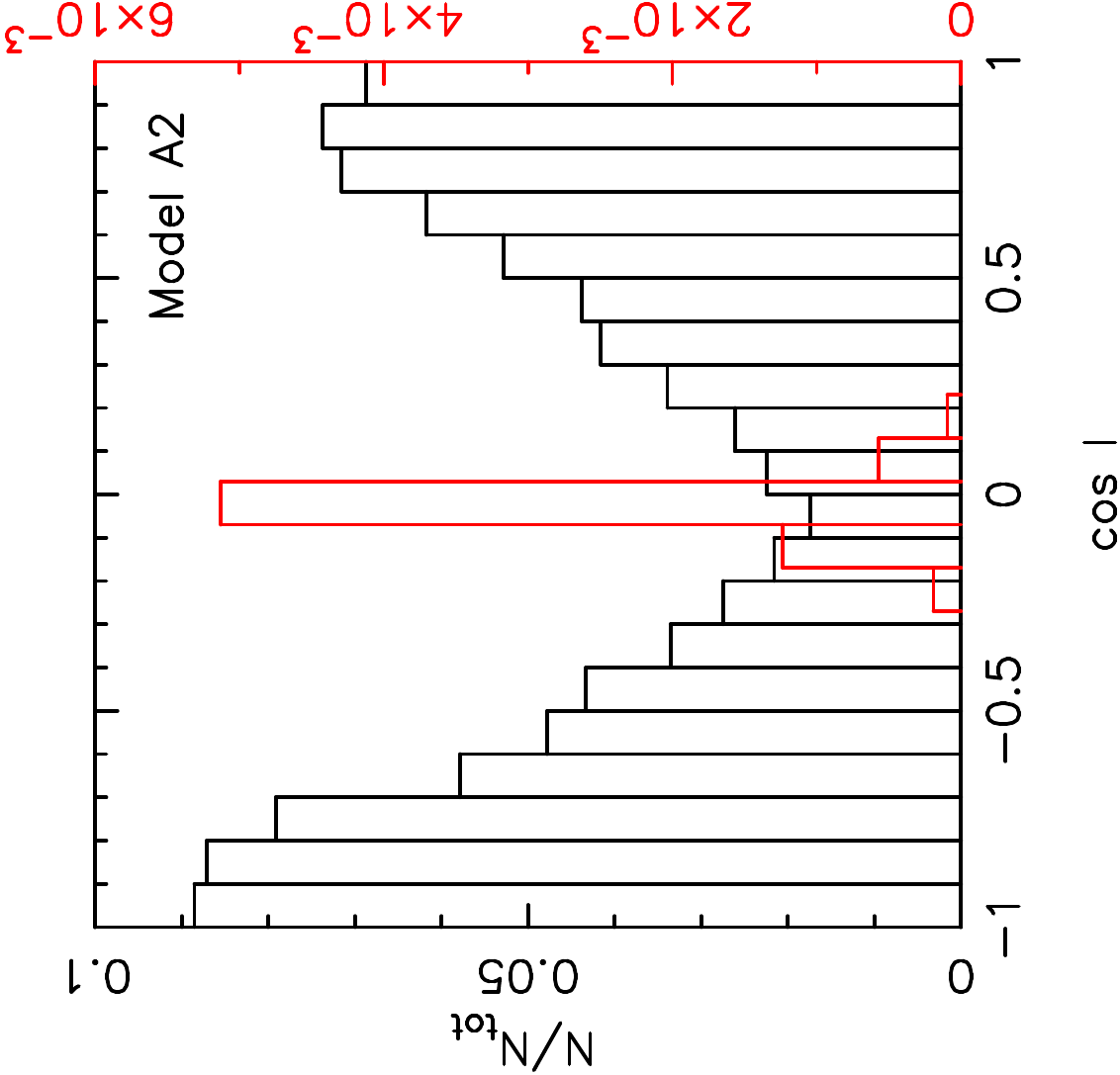}\\
\raisebox{-2.4mm}{ \includegraphics[angle=270,width=2.4in]{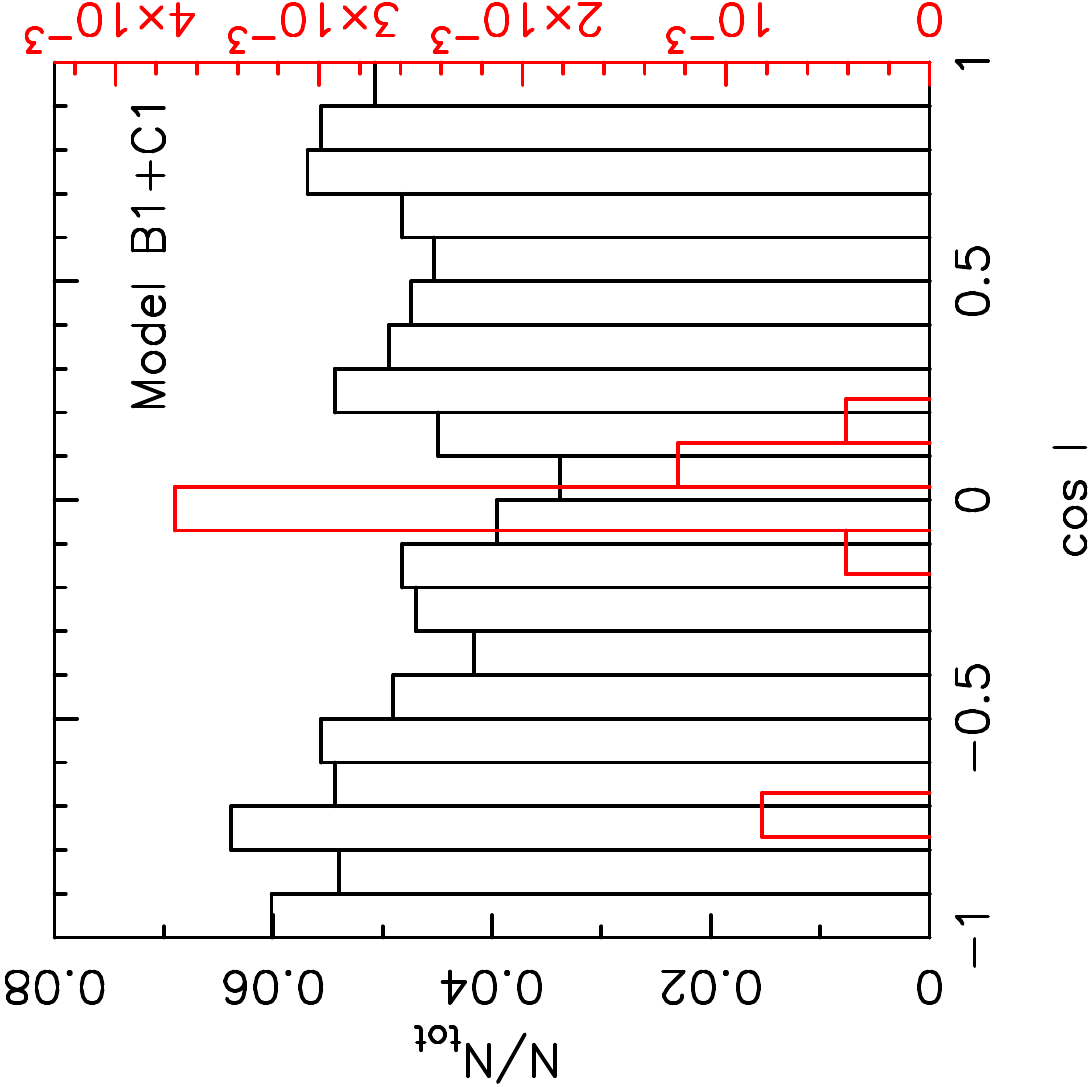}  }
 \includegraphics[angle=270,width=2.4in]{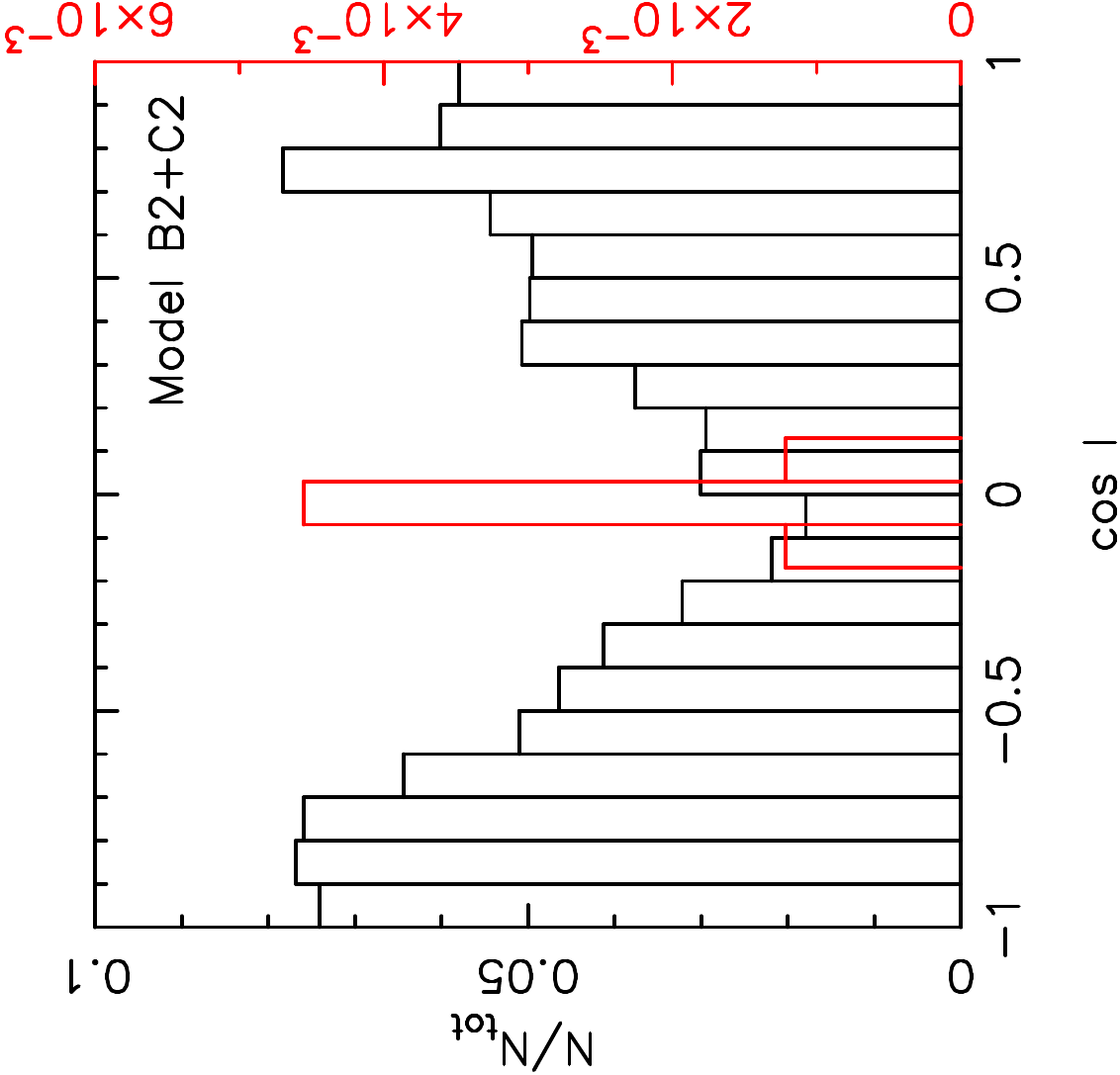}
   \caption{Distribution of relative inclination of outer to inner orbit for the BH triples
   formed in our models according to the secular population synthesis code (black histograms), and for the triples 
   which produce 
   BH mergers (red histograms). 
   The number of systems 
are normalized by the 
  total number of stable BH triples formed. 
   This plot shows   that the initial distribution of $I$ is not isotropic
and that   most merging BH binaries are produced in BH triples with initially high mutual inclinations.
}\label{incl}
   \end{center}
\end{figure*}

\begin{figure*}
\begin{center}
 \includegraphics[angle=270,width=2.3in]{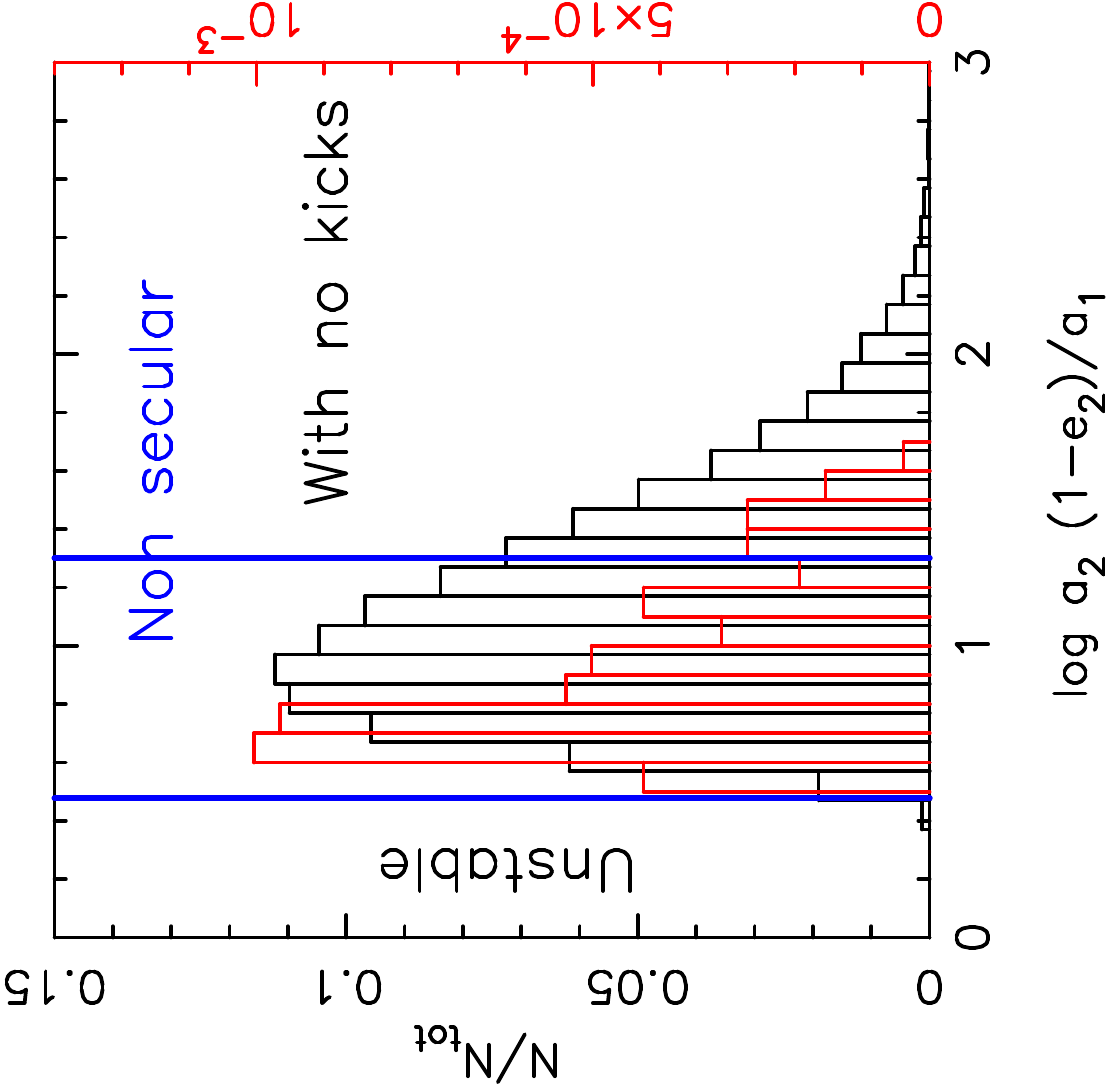} 
\raisebox{+.2mm}{ \includegraphics[angle=270,width=2.3in]{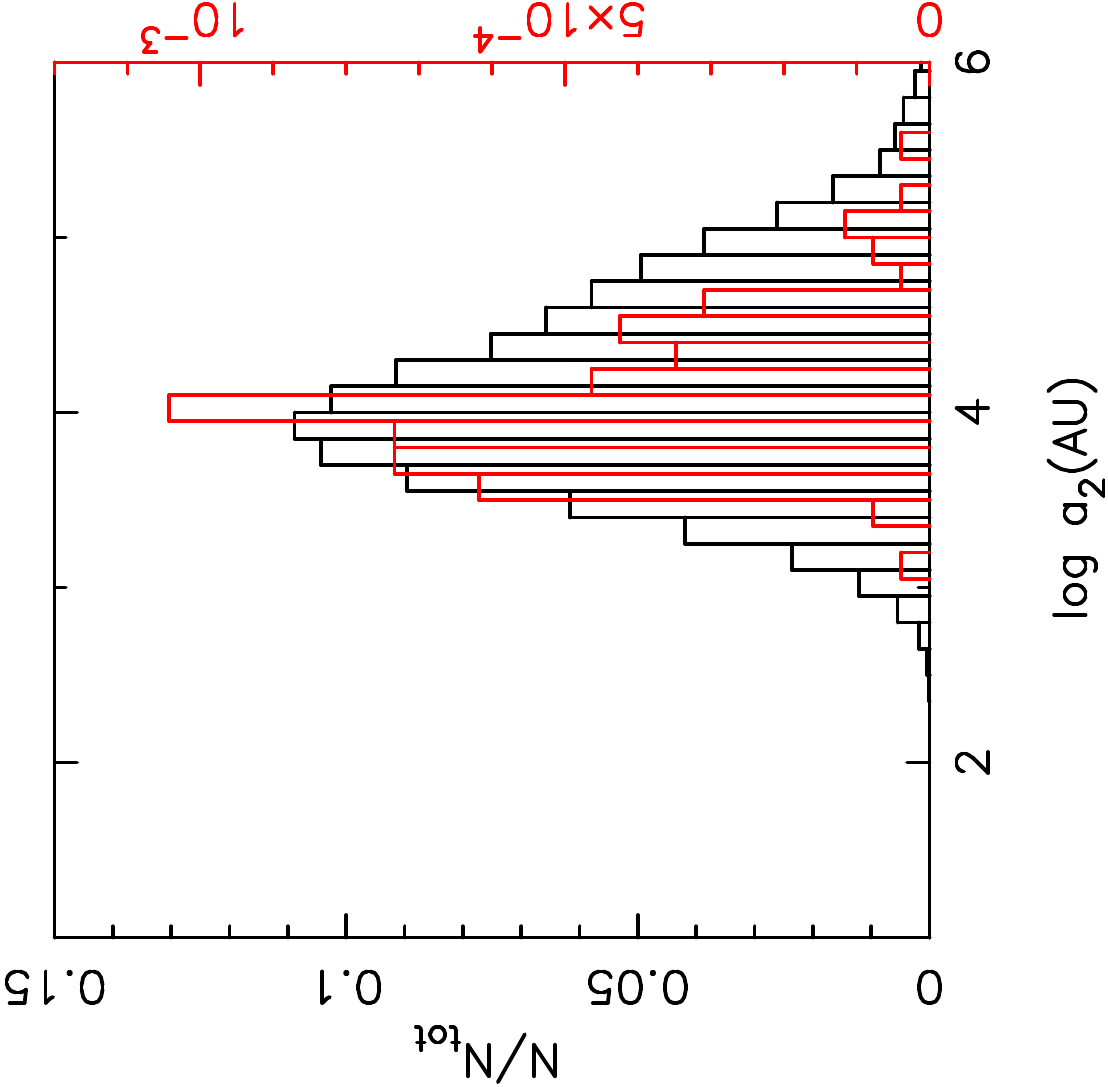}}
\raisebox{-0.mm}{  \includegraphics[angle=270,width=2.3in]{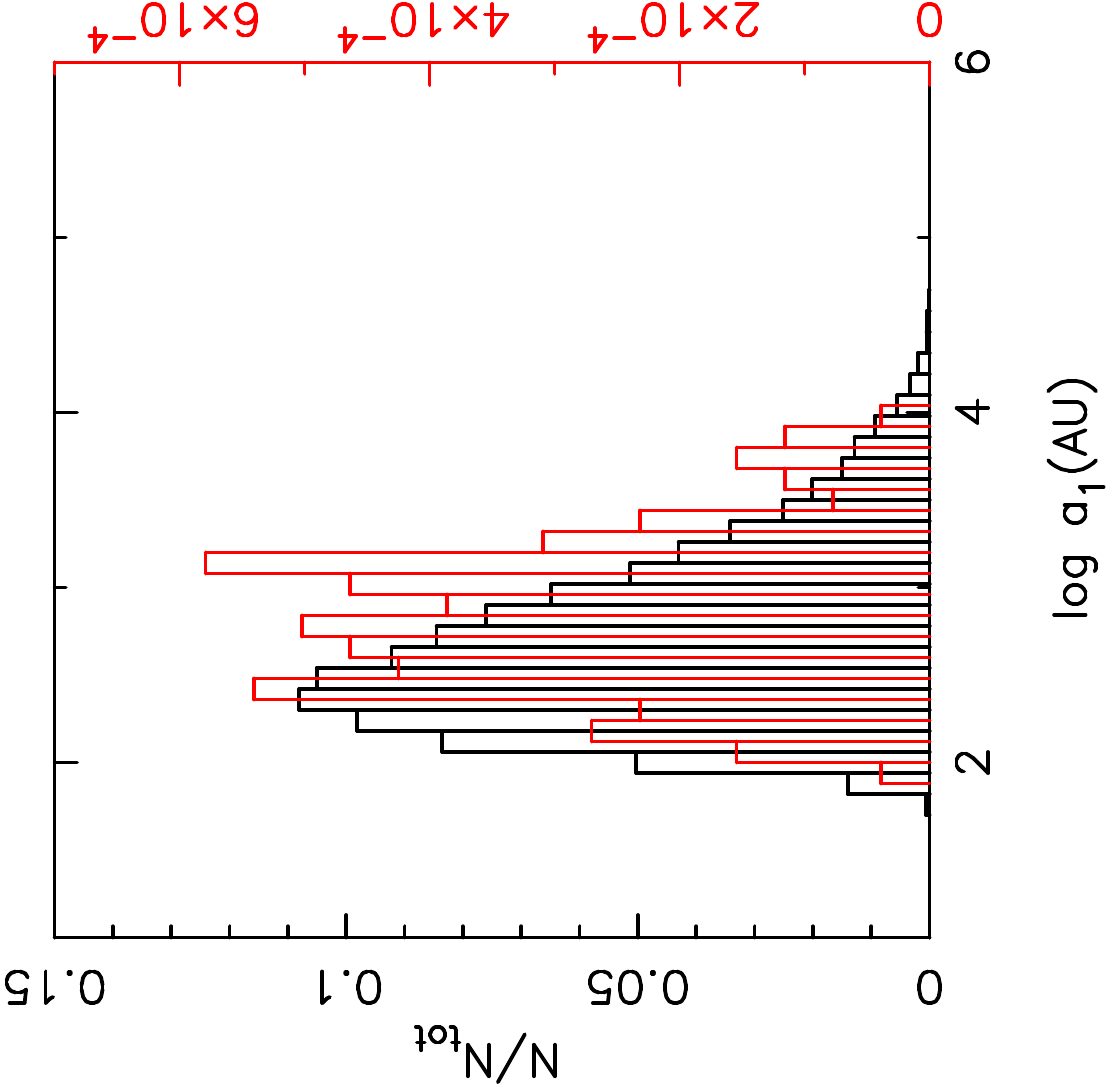}}\\
 \includegraphics[angle=270,width=2.3in]{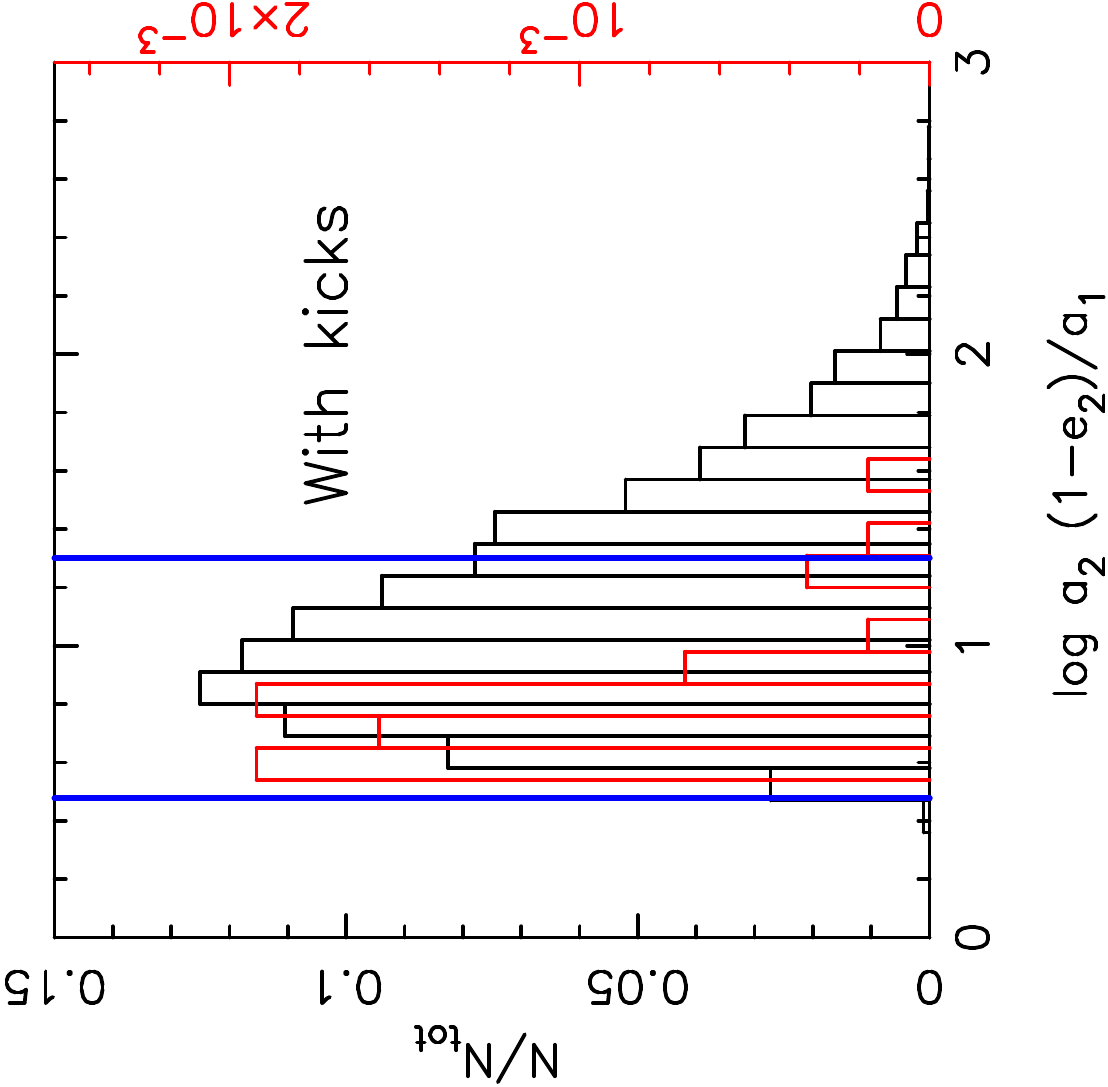}
\raisebox{+2.2mm}{\includegraphics[angle=270,width=2.3in]{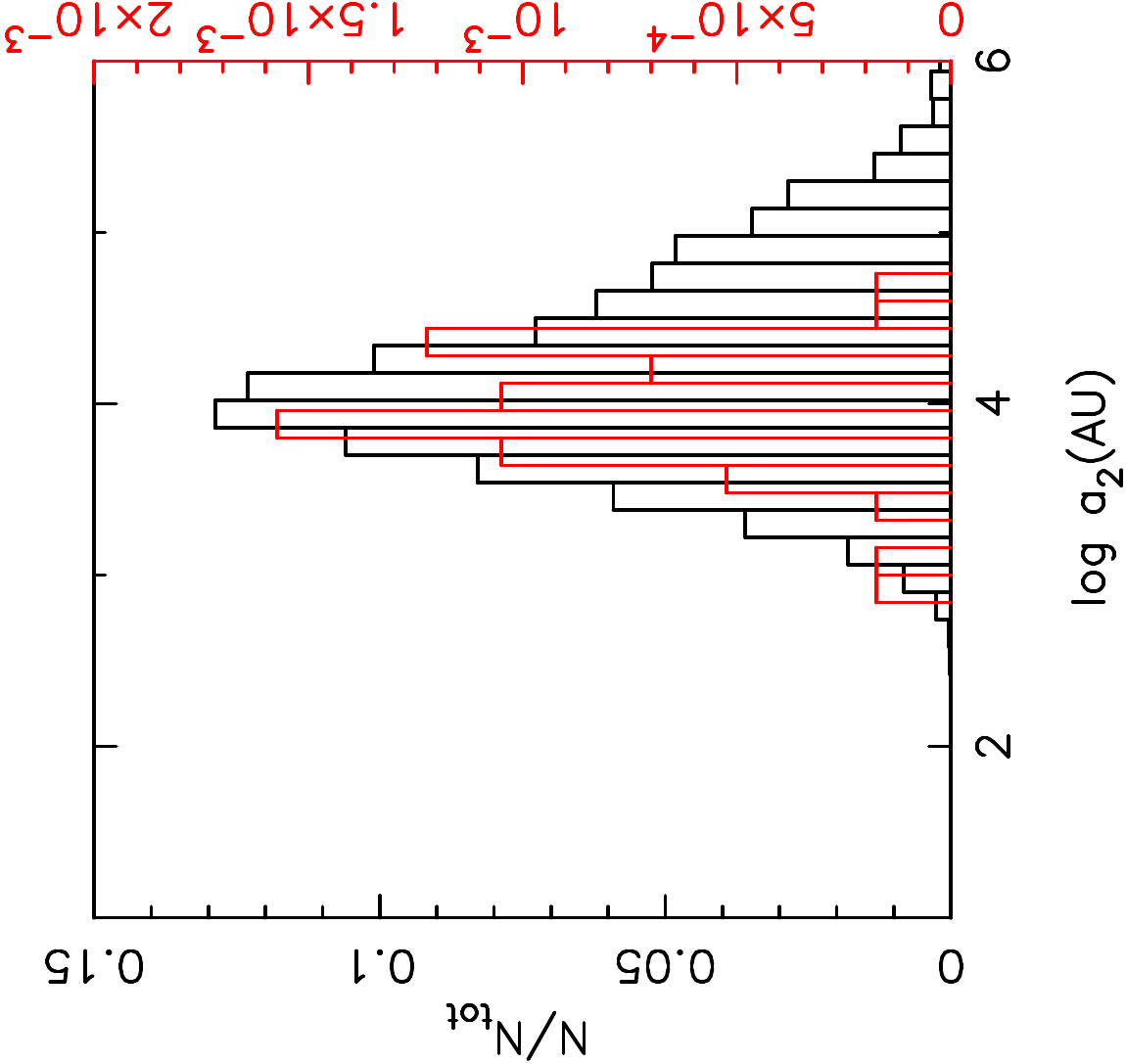}}
{  \includegraphics[angle=270,width=2.3in]{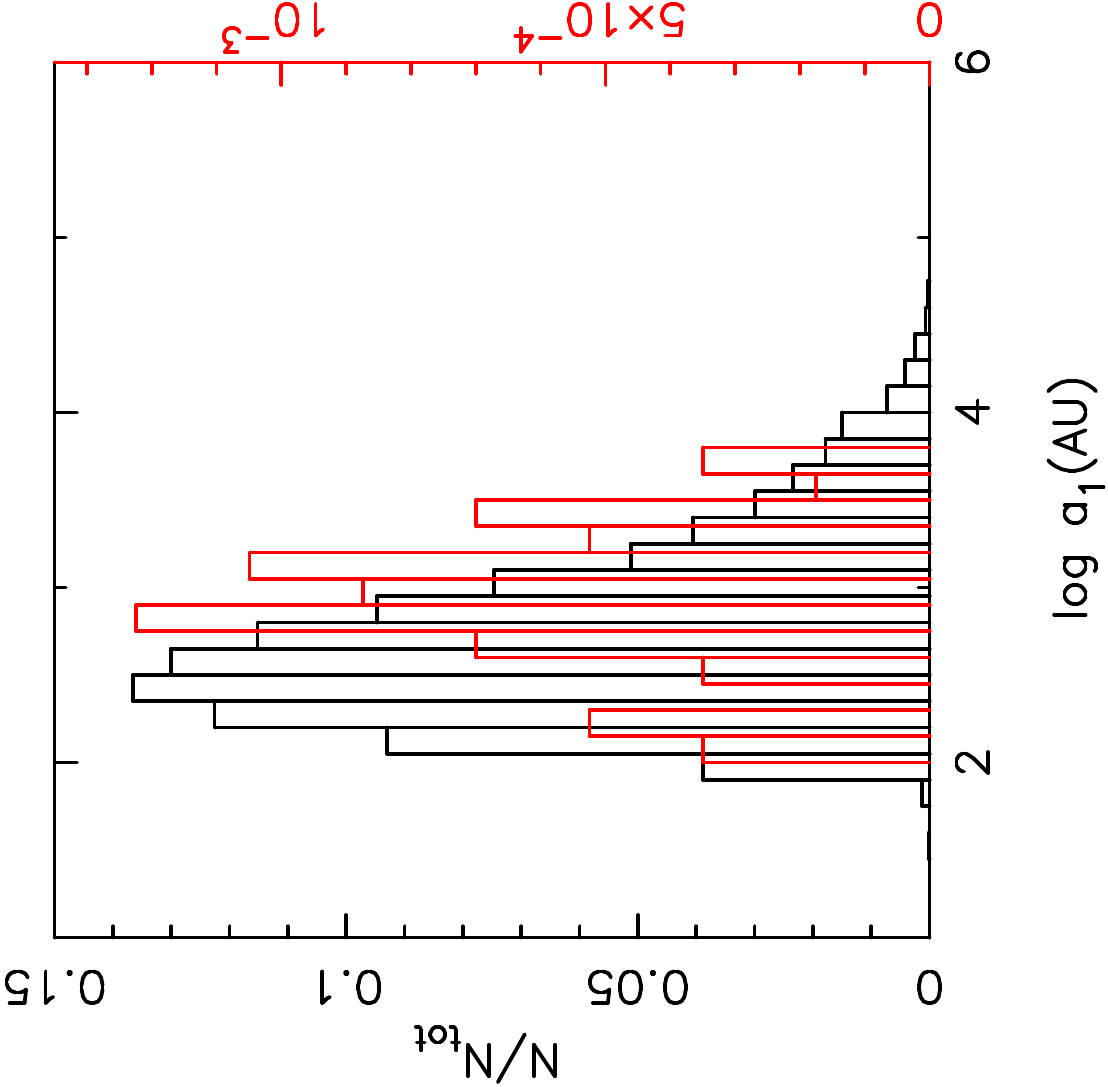}}\\
  \caption{ 
  Distribution of orbital separations of the stable BH triples formed in  
  the secular population synthesis simulations (black)
    and those that lead to a merging BH binary (red).
  Left panels give the
   distribution of the minimum distance of the outer BH to the inner binary,
  divided by the inner BH binary semi-major axes; the middle panels show the
  distribution of  the outer orbit semi-major axes; the right panels show the distribution of 
the inner binary semi-major axis. These distributions have been normalized to the total number of BH triples formed.
 In the left panels,  the BH triples that are within 
  the blue vertical  lines and achieve  an eccentricity $e_1\gtrsim e_{\rm crit}$ enter the non-secular dynamical regime defined by Equation\ (\ref{aa})
  before GW radiation becomes important to their evolution.  
  Most triples that produce BH mergers 
 are expected to  evolve in the non-secular regime where the standard octupole secular equations  of motion
   \citep[e.g.,][]{2013MNRAS.431.2155N,2016ARA&A..54..441N}
  are not valid. The right panels show that BH binaries formed through the triple
  channel are driven to a merger from distances $ 10^2\lesssim a_1\lesssim 10^3\rm AU$ and have companions at 
  $10^3\lesssim a_2\lesssim 10^4\rm AU$.}\label{sepplot}
\end{center}
\end{figure*}

\begin{figure*}
\begin{center}
 \includegraphics[angle=270,width=2.3in]{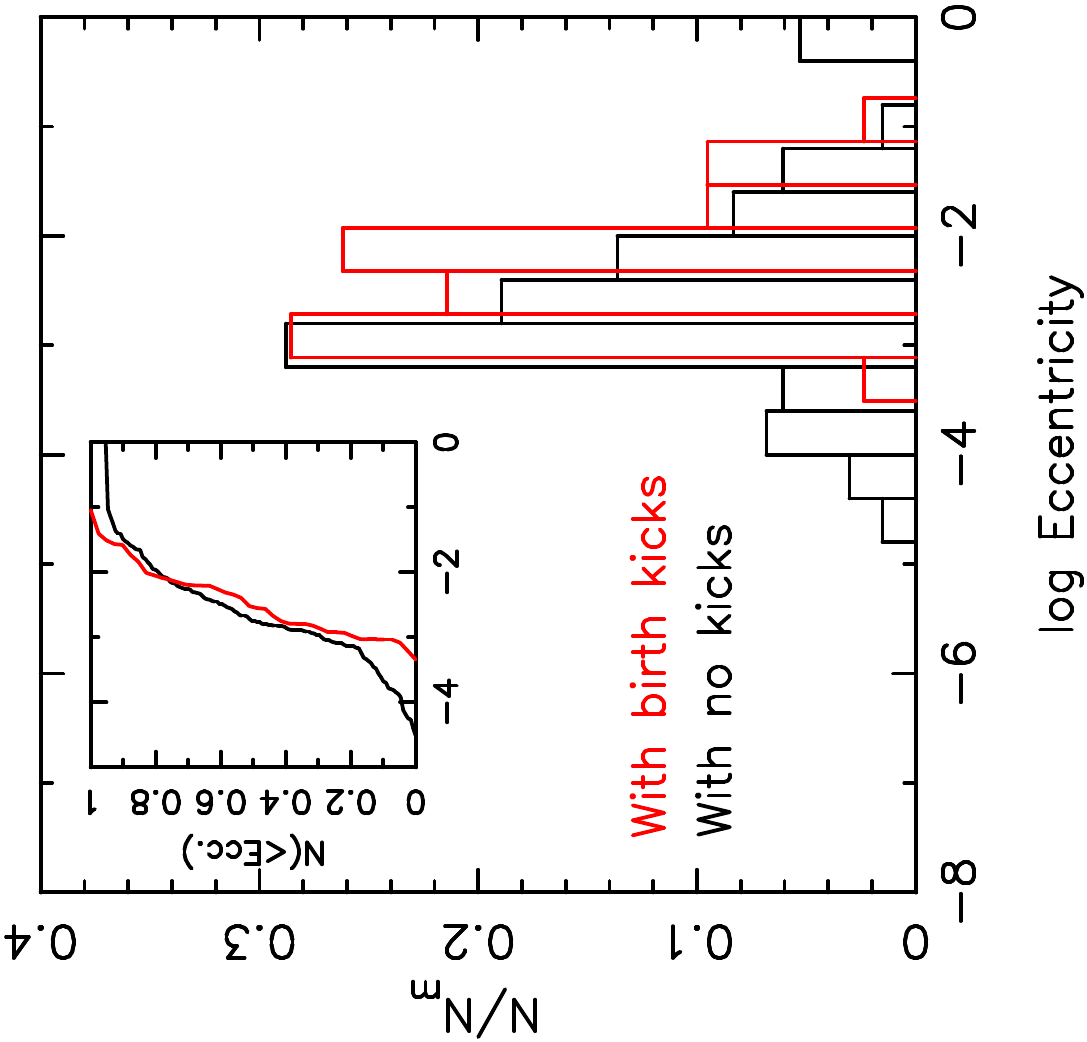}  
 \includegraphics[angle=270,width=2.165in]{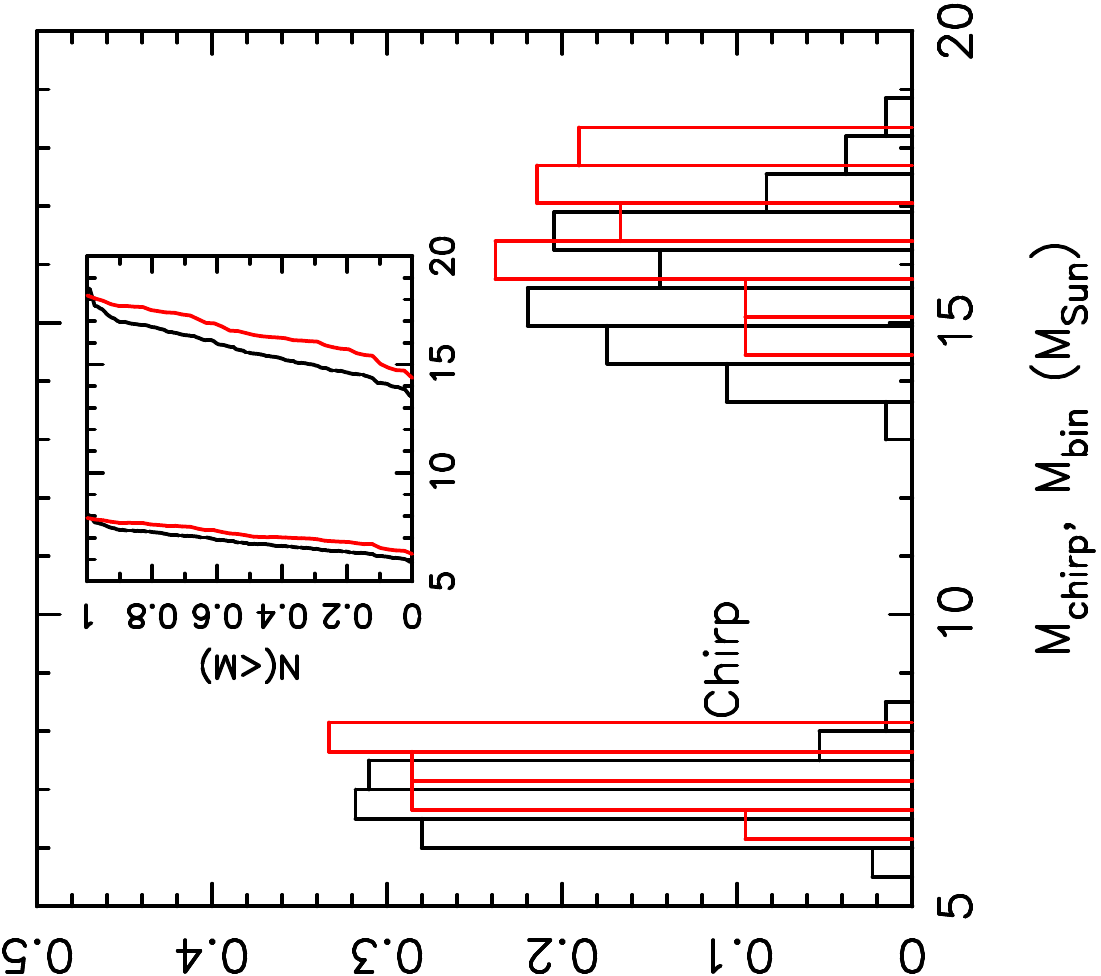}
 \raisebox{+.8mm}{ \includegraphics[angle=270,width=2.12in]{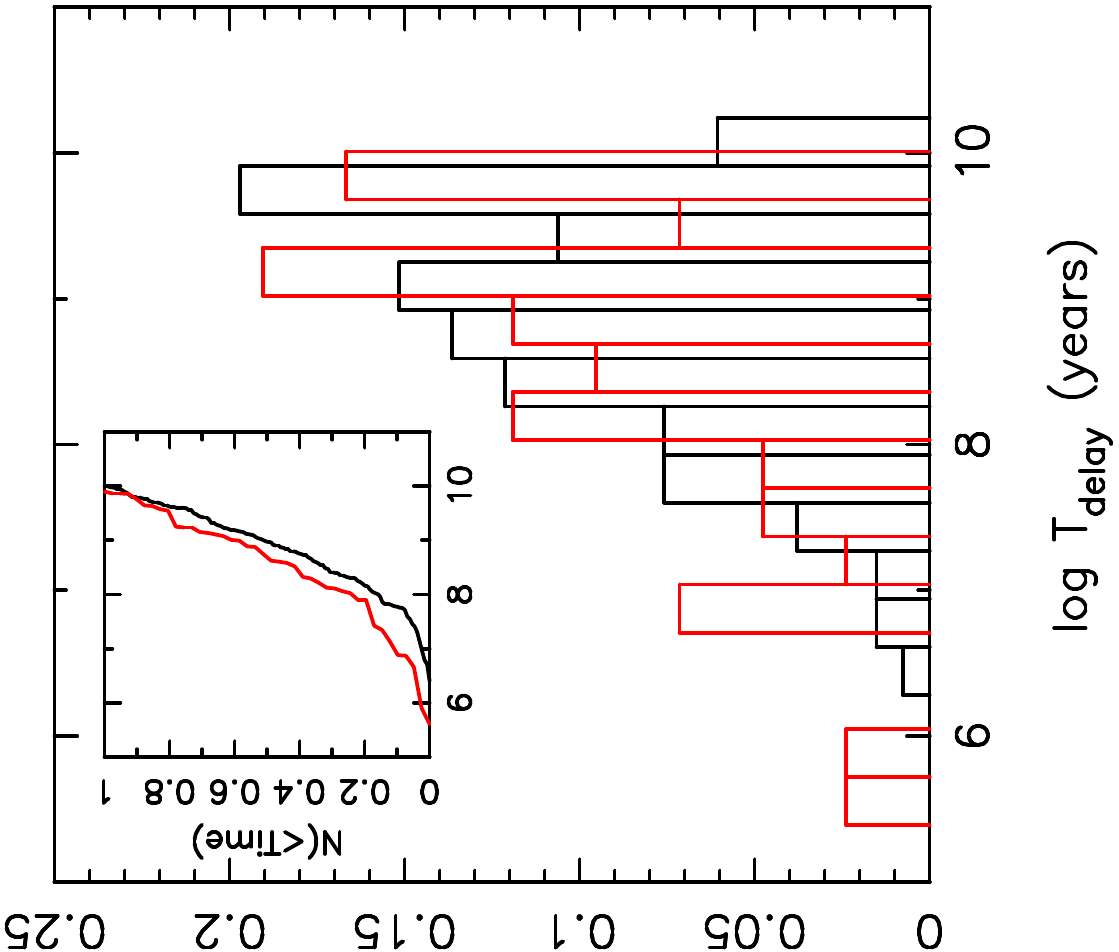}}
  \caption{Properties of merging BH binaries. Distributions are normalized by the 
  total number of mergers. The left panel
  gives the distribution of eccentricities at the moment the binaries first enter the aLIGO frequency 
  band. The middle panel shows the distribution of total mass, and chirp mass of 
  the merging BHs. The right panel gives their time delay distribution, i.e., the time from formation to coalescence.
Red histograms and curves are for models B1, B2, C1 and C2 combined; black is for models A1 and A2.
  }\label{eccen}
\end{center}
\end{figure*}

 \begin{table*}
  \caption{Results of the population synthesis models of massive triple stars.}
  \centering
\begin{tabular}{llllllllllllll}
\hline 
\hline
Model& natal kicks & $a_{2;\rm max}$ &  $m_{3;\rm min}$ &$N_{\rm sim}$ & 
 fraction& fraction &fraction &
$\epsilon_\mathrm{3{BH}}$ &
$\epsilon_\mathrm{merge}$ & $\Gamma $ \\
& & $\rm (10^3AU)$ &  $\rm (M_\odot)$ & & 
disrupted  
& mass transfer&  dyn. unstable&
 &   &$(\rm  Gpc^{-3}yr^{-1})$ \\
  \hline 
A1& 0           & $20 $ & 0.1 & 50k & 0.31 & 0.27 & 0.19  &0.18 &$6\times 10^{-3}$& 1.3 \\
A2&0            &$ 2$ & 22 & 50k &  0.14 &  0.43 & 0.22 &0.21   &$7\times 10^{-3}$&  1.2 \\
B1 & Hobbs &  $20$  & 0.1& 25k& 0.56  & 0.26& 0.12 &0.06   &$5\times 10^{-3}$ &  0.4 \\
B2 & Hobbs &  $2$& 22 &  25k & 0.36   & 0.41& 0.17&0.06   &$1\times 10^{-2}$ & 0.5 \\
C1 & Arzoumanian & $20$ & 0.1 & 25k & 0.56&  0.26&  0.13 &0.05  & $8\times 10^{-3}$ & 0.5\\
C2& Arzoumanian& $2$& 22 &25k& 0.35&  0.42&  0.17 &0.06 &  $5\times 10^{-3}$  & 0.3\\
%\hline
%A1-r& 0           & $20 $ & 0.1 & 75k & 0.31 & 0.27 &  0.23 &$7\times 10^{-3}$& 1.4 \\
%A2-r&0            &$ 2$ & 22 & 75k &  0.14 &  0.43 & 0.21   &$1.5\times 10^{-2}$&  2.1 \\
  \hline \hline
\end{tabular}
{\\ Models differ by their natal kick velocity distribution,
adopted 
  maximum value of separation, $a_{2,\rm max}$, and minimum  mass, $m_{\rm 3; min}$, of the tertiary 
  star. $N_{\rm sim}$ is the total number of stellar triples
  that were evolved for each model; 
   $\epsilon_\mathrm{3{BH}}$  is 
  the fraction of systems that produce stable BH triples;
  $\epsilon_\mathrm{merge}$ is the fraction of the stable BH triples
  that lead to the merger of a BH binary; and $\Gamma$ is the inferred BH binary merger rate.
  }\label{t1}
\end{table*}

 \section{Results}\label{results}
   In this section, we describe the key physical properties  and merger rates 
   of the BH binaries/triples in our models.
     In total we observe 177 BH mergers among the 200,000 systems we 
evolved.
 We note that  systems other than BH triples,
 also  produce mergers in our simulations. Main-sequence stars that do not evolve to a compact object within one Hubble time
 are unlikely to result in efficient LK cycles due to their lower mass. Accordingly, we did not find 
 any mergers for systems in which the tertiary was still on the main-sequence  at the end of the simulation.
  In our models  with low
mass tertiaries (i.e., models A1, B1 and C1),
 we  found that 8 out of a total of 75 mergers  were 
produced in triples in which the tertiary was not a BH.
In 3 of these systems the tertiary companion became a neutron star  
by the end of the simulation, in 3 the tertiary became a CO white dwarf, 
and in the remaining 2 an ONe white dwarf.

Figure\ \ref{example} shows the results of
a three-body direct integration of
one BH triple system that leads to the formation of a BH merger.
The secular exchanges of angular momentum (but not energy) among
the inner binary and the outer BH  induce  large  fluctuations  in  the inner binary  eccentricity
and  inclination.   The  orbit  of  the  inner  binary,  starting  from an initial  value  of $e_1=0.65$ 
  and $I=93^{\circ}.8$,  
diffuses   to $1-e_1\approx 1\times 10^{-4}$ by $\approx 3\times 10^6\rm yr$.
 We find that the maximum orbital eccentricity of the inner binary
undergoes a random walk to most of the phase space allowed by the total energy
 and angular momentum of the system. 
 During the maximum of a LK oscillation 
 the binary enters the  non-secular regime defined by Equation\ (\ref{aa}).
 In this region the inner BHs can be driven to a merger before  general relativistic effects
suppress the secular forcing. At the end of the integration, GW radiation starts to dominate the binary evolution. 
Subsequently, the BH binary starts to circularize, decouples from the tertiary companion, and 
finally enters the $10\rm Hz$ aLIGO frequency band with $e_1= 0.4$ and 
$a_1=1.5\times 10^{-5}\rm AU$.

  \subsection{Properties}

Table 1 gives the fraction of systems that undergo mass-transfer, that are disrupted 
due to supernova kicks, and the fraction of stable BH triples, $\epsilon_\mathrm{3{BH}}$, that are formed in our models.
The rest of the systems became dynamically unstable  \citep{2001MNRAS.321..398M} during their evolution due to e.g. stellar wind mass
loss  \citep{2012ApJ...760...99P} and supernova kicks.
Simulations with the smaller
range of masses/orbits (models A2, B2 and C2) show:
fewer systems that are disrupted due to the supernova kicks, more systems that experience mass transfer,
and an almost unchanged fraction of stable BH triples.
Compared to models with no kicks, the simulations with non-zero birth kicks (models B1, B2 and C1, C2) show:
more systems that are disrupted due to the supernova kicks, and, consequently, fewer 
stable BH triples formed.

    In Figure\ \ref{incl}, we show the distribution of the inclination between inner and outer orbit
  of the initial BH triples (black histograms), and of the subset of these that
  produce BH mergers (red histograms). 
  Figure\ \ref{incl}  shows that essentially all merging binaries are formed 
 in triples with high inclination and in the range $70^\circ\lesssim I \lesssim110^\circ$.
  The initial BH triples are formed with
  a distribution which is clearly anisotropic, with a deficit of orbits near $\cos I= 0$.
  This is a consequence  of many initially highly inclined systems that
  due to the LK mechanism merge or undergo a  phase  of mass transfer before the 
  three BHs are formed. This effect, 
  similarly occurring in the evolution of lower-mass triples \citep{2013MNRAS.430.2262H},
  reduces the number of BH triple systems with 
 high inclination which can lead to a BH binary merger,  lowering the  overall BH merger rate
compared to what we would obtain by assuming an initially random distribution of 
inclinations \citep[e.g.,][]{2016arXiv160807642S}.

Figure\ \ref{incl}  shows  that the $I$ distribution of BH triples with initially larger 
semi-major axis is closer to isotropic. The reason for this is that systems where the tertiary is
at larger separation are less likely to merge during their main-sequence evolution
given their longer LK timescale 
(see {Equation}~[\ref{tkl}]) and that 
the $e$-oscillations are more strongly quenched by precession of the periapsis due to relativity  or  due to tidal bulges 
for larger values of $a_2/a_1$   \citep[e.g.,][]{2002ApJ...578..775B,2007ApJ...669.1298F}. 

Figure\ \ref{sepplot} gives the distribution of
the semi-major axes $a_1$ and $a_2$ at the moment the triple BHs are formed and 
for the subset of the BH triples that lead to  merging binaries. These plots 
show that our mergers are produced in BH triples  with $a_1\gtrsim 100\rm AU$, and $ a_2\gtrsim 1000\rm AU$.
The left panels of Figure\ \ref{sepplot} give the distribution of 
the ratio $a_2(1-e_2)/a_1$. This latter quantity
 parametrizes how well the dynamics of the system can be described
by the standard secular equations of motion.
The non-secular region  in Figure\ \ref{sepplot} 
was computed from
Equation\ (\ref{aa})  by requiring the inner BH binary to reach an eccentricity 
large enough for GW radiation to dominate its evolution.
One  finds that the inner  binary  can  enter the non-secular regime
 if \citep{2016ApJ...816...65A}:
\begin{equation} \label{nss}
{a_2 (1-e_2)\over a_1}\lesssim {2.5}  \left(m_3\over m_{\rm 1}+m_2\right)^{1/3} 
\left( a\over D_{\rm diss}\right)^{1/6},
\end{equation}
with $D_{\rm diss}$ the typical dissipation scale.
Most of our binaries are driven to a merger from an initial distance $a_1\gtrsim \rm 100\rm AU$. 
Adopting  $D_{\rm diss}=10^9\rm cm$ as a conservative dissipation scale 
and assuming equal-mass components, 
we find that
the condition  Equation\ (\ref{aa})  is met before  GW radiation dominates the evolution of the inner binary
 if  ${a_2 (1-e_2)/a_1}\lesssim 20$.
We stress  that not all triples that are within  the blue vertical lines  in Figure\ \ref{sepplot}
 enter the non-secular regime, but only those
that achieve  an eccentricity $e_1\gtrsim e_{\rm crit}$. 
The red histograms in Figure\ \ref{sepplot} represent the binaries that merge in our models.
Evidently, most merging binaries in our simulations are expected to evolve through the non-secular dynamical regime
defined by Equation\ (\ref{aa}); this happened for 132  of the 177 total mergers  we found.
 
The left panel of Figure\ \ref{eccen} displays the distribution of eccentricities 
at the moment  the peak GW frequency of the binaries becomes larger than the 
$10\rm Hz$ aLIGO frequency band.
Eccentric binaries emit a GW signal with a broad spectrum of
frequencies; we compute a proxy for the GW frequency  of our merging
binaries as the frequency corresponding to the 
harmonic which leads to the maximal emission of GW
radiation \citep{2003ApJ...598..419W}:
\begin{equation}\label{fgw}
f_{\rm GW}={\sqrt{G(m_{\rm 1}+m_2)}\over \pi}
{ \left(1+e_1\right)^{1.195} \over \left[a_1\left( 1-e_1^2\right)\right]^{1.5}} \ .
\end{equation}
About $3\%$ of all merging BH binaries in our models enter the $10\rm Hz$ window 
with an extremely high eccentricity $(1-e_1) \lesssim 10^{-6}$, while 
$\approx 90 \%$ of them have eccentricities $\lesssim 0 .1$ at $10\rm Hz$. 

\begin{figure}
\begin{center}
 \includegraphics[angle=270,width=2.8in]{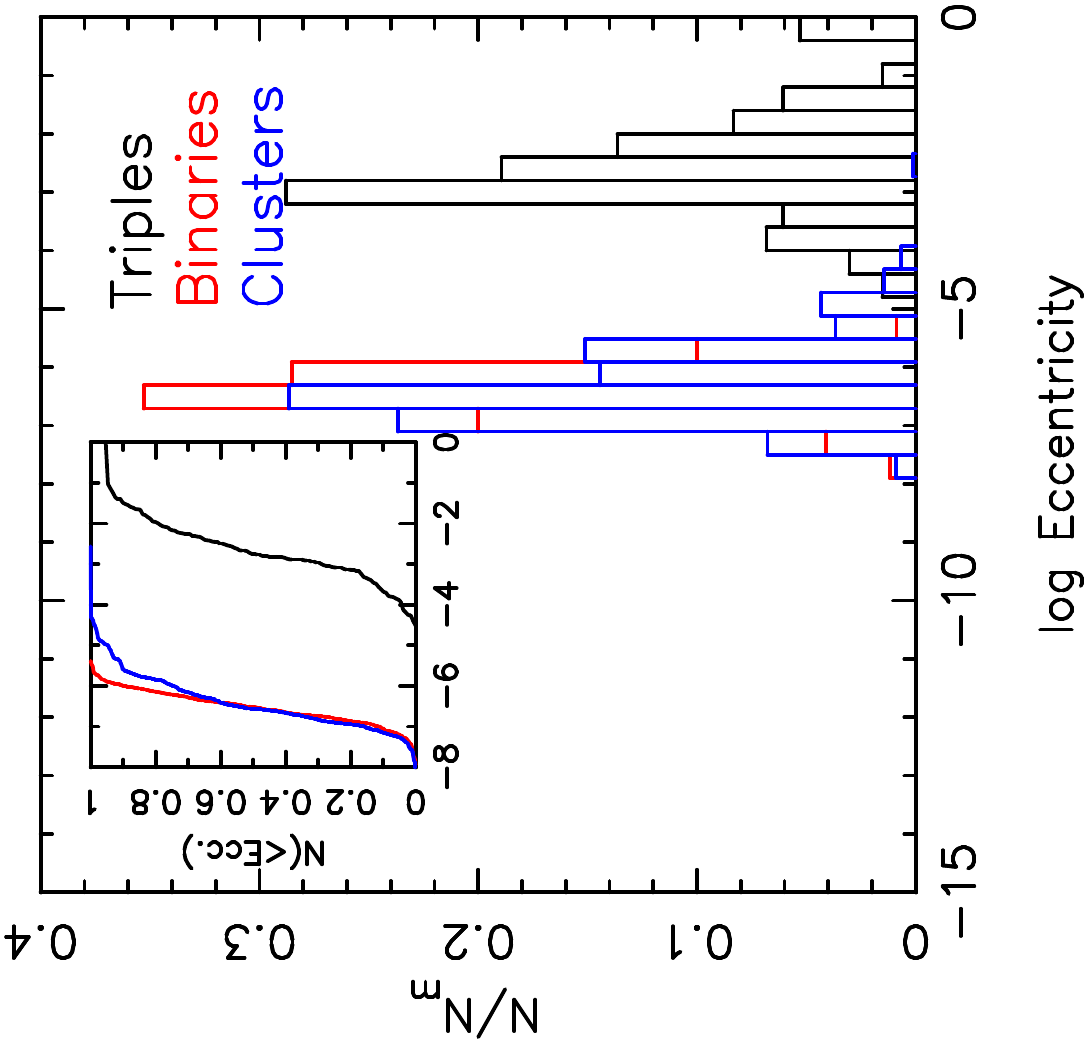}  
  \caption{Distribution of eccentricities at the moment the  BH binaries enter
  the aLIGO frequency band ($10\rm Hz$) for
  mergers produced by dynamical interactions in 
  dense star clusters,
  massive binary stars,
   and massive triples (models A1+A2).
  Binaries formed in triples have much larger eccentricities than
  those formed through  other  channels.}\label{ecc-comp}
\end{center}
\end{figure}

In Figure\  \ref{ecc-comp},
we compare the eccentricity distribution at $f_{\rm GW}=10\rm Hz$ of our merging BH
binaries to those formed in star clusters and from field binaries. These two latter distributions were taken from Figure 3 
of  \citet{2016ApJ...830L..18B}.
While the field and cluster models predict similar eccentricity
distributions \citep[e.g.,][]{2017MNRAS.465.4375N},
 the eccentricities of merging BH binaries from field triples appear to be uniquely biased towards high values.
 We conclude that eccentricity measurements alone could be 
potentially used  to discriminate  binaries formed through the evolution of isolated massive triple stars.
However, the high eccentricities found in our models
also imply that a fraction of these binaries could emit their maximum power 
at frequencies much higher than the frequency 
window of the
planned Laser Interferometer Space Antenna (LISA; $f_{\rm GW}\approx [10^{-3},0.1] \rm\ Hz$).
 As argued in \citet{2017arXiv170208479C},  binaries that enter the aLIGO band 
 with $e_1\gtrsim 5\times10^{-3}$ could be harder to detect with instruments like LISA.
Of the 177 merging binaries in our simulations, 69 (137) have eccentricities  higher than $5\times10^{-3}$  ($10^{-3}$)
and  could therefore be harder to detect at lower frequencies.

In the middle panel of  Figure\ \ref{eccen}, we show the distribution of the total and chirp mass of the 
merging binaries, where this latter is defined as $M_{\rm chirp}=(m_1m_2)^{3/5}/(m_1+m_2)^{1/5} $. 
Our merging binaries have total masses in the range [$13\rm M_{\odot} $,~$20\rm M_{\odot}$], and 
a chirp mass distribution that peaks around $\approx 7M_{\odot}$.
The merging binaries in models with a non-zero natal kick 
appear to have larger masses when compared to the models  without birth kicks. 
This is expected: given our assumption of momentum-conserving kicks,
higher mass BHs receive in average lower velocity kicks  and are therefore
more likely to be retained in triples and merge.

The right panel of Figure\ \ref{eccen} shows the time delay distribution,
where $T_{\rm delay}$ is the time from formation of the BH triple to the merger of the inner BH binary.
 In models with no
birth kicks, about $\approx 50\%$ of all merging binaries have delay times  $\lesssim 1\rm Gyr$, while 
 this percentage increases to $\approx 60 \%$ in models with non-zero natal kicks. 
Essentially all merging binaries  have delay times larger than $\approx 10^{6}\rm yr$.
 Note that for a $22M_\odot$ star the time from formation of the star  to the formation of the
 BH is $\lesssim 8\rm\ Myr$.

 \subsection{Merger rates}
 In order to compute the merger rate of binary BHs, we follow the procedure described 
 in \citet{2016arXiv160807642S}.
The number of stars formed per unit mass 
is given by:
 \begin{equation}
 N_{\star}(m)dm=5.4 \times 10^6 m^{-2.3} \rm  Gpc^{-3} yr^{-1}.
  \end{equation}
  Adopting a constant star formation rate per comoving volume 
 unit, the merger rate of binary BHs is then:
   \begin{equation}\label{rt}
\Gamma\approx N_{22,100} \epsilon_{\rm prog}\epsilon_{\rm p-space} \epsilon_\mathrm{3{BH}}  \epsilon_\mathrm{merge},
  \end{equation}
 where $N_{22,100}=6\times 10^4 \rm  Gpc^{-3} yr^{-1}$ is the number of stars formed with mass between $22$
 and $100$ solar masses.

 The quantity $\epsilon_{\rm prog}$ in Equation (\ref{rt})
is the ratio of  BH triple progenitors 
 to total BH progenitors.
 As in  \citet{2016arXiv160807642S}, we assume the following:
 $19\%$ of systems with at least one BH progenitor are 
 single systems, $56\%$ are binaries, and 
$25\%$ are triples. These percentages  are consistent with 
\citet{2014ApJS..215...15S} who found these fractions of multiplicity for the  O-stars in their sample.
Given our mass distributions,  for models A2, B2 and C2, we find a fraction  
  $\epsilon_{\rm prog}=0.04$  
  of triple progenitors with $(m_1,~m_2,~m_3)>22 M_{\odot}$, 
  and for models A1, B1 and C1 we find a fraction 
$\epsilon_{\rm prog}=0.06$ 
 of  triple progenitors in which the inner binary components have
 ZAMS mass $(m_1,~m_2)>22 M_{\odot}$.

The quantity $\epsilon_{\rm p-space}$ is  the fraction of parameter space that we are simulating 
relative to the full parameter space for massive triples that is covered by observations.
This fraction is $\epsilon_{\rm p-space}\approx 0.35$ in our models and takes into account the
fact that we are only simulating systems with $a_1(1-e_1^2) \gtrsim 11\rm\ AU$
initially.
In order to estimate $\epsilon_{\rm p-space}$ we have
assumed a maximum separation for the outer orbit of
$6000\rm\ AU$ as this is approximately  the  distance within which 
the companions to the  massive stars in the sample of \citet{2014ApJS..215...15S}
 are resolved or spectroscopically identified.

Finally,  $\epsilon_\mathrm{merge}$
is the fraction of dynamically stable BH triples formed in our models
which produce a BH merger. Roughly speaking, this fraction, $\approx 1\%$, is
 independent  on the assumed distribution of 
natal kick velocities. This latter result can appear quite surprising:
 the  change  in  linear  momentum
instantaneously  imparted  to  the  exploding  stars   alters
the orientation of orbital planes subsequent to BH formation
and could result  in a larger number of BH triples that are formed with initially high inclination. Consequently, one would expect
 a larger fraction of merging binaries   
in the non-zero kick models \citep[e.g.,][]{2016arXiv160807642S}.
Contrary to this expectation, our simulations suggest that this effect has a small 
impact on the resulting BH binary merger rate.
We note also  that $\epsilon_\mathrm{merge}$ appears to be approximately
unaffected  by the choice we make for the upper limit on the separation of the outer
orbit.

 Table 1 gives the results of our calculations. We estimate the BH merger rate in 
 isolated triple systems in the field to  be at most $\approx 1\rm\ Gpc^{-3} yr^{-1}$.
 
 Some of our models can be directly compared to the results of 
  \citet{2016arXiv160807642S}. For example, similar to our models A1 and A2, the latter authors 
  also consider models where the BHs receive no natal kicks (see their Table 2).
    Their zero-kick models produce  a merger rate of $\approx 6\rm\ Gpc^{-3} yr^{-1}$, which 
   is about six times larger than the merger rate  inferred from our simulations. 
One reason for  the discrepancy in the rate estimates is that
     \citet{2016arXiv160807642S} 
   assume zero Blaauw kick \citep{1961BAN....15..265B},  which increases the chance that a triple in their zero-kick models 
will survive the formation of a BH, leading to  higher merger rates. 
  Moreover,  \citet{2016arXiv160807642S} assume that
BH triples are formed with initially random inclinations. 
However, in many of the highly inclined triples  in our models, the two inner objects merge 
early during their main-sequence evolution. 
We note also that  \citet{2016arXiv160807642S} used a somewhat different model for the period and mass distributions,
and a shorter maximum simulation time. These differences with our models will also affect the merger rates, 
but not as significantly as the effects due to the more realistic inclination and kick distributions used in our simulations.

    \citet{2016arXiv160807642S} also consider models with a non-zero natal  kick velocity. 
In one of their models, they adopt a Gaussian kick velocity distribution with $\sigma \approx 40\rm km\ s^{-1}$
 which results in a  merger   rate of $\approx 0.14\rm\ Gpc^{-3} yr^{-1}$ (see their Table 2). 
 This appears to be a few times smaller
  than the rate inferred from our  models B1, B2 and C1, C2.
 We believe that the main reason for our higher merger rates in this case is that
  we implemented a zero kick for the very massive stars \citep{2001ApJ...554..548F},
which  increases the number of systems that remain bound after BH formation.

   \section{Additional Considerations}\label{ac}
        
 \subsection{Consequences  of non-secular dynamics}
As shown above in Figure\ {\ref{sepplot}},  the merging binaries 
in our models evolve through a non-secular dynamical phase where the standard 
secular perturbation theory is expected to break down. This has important consequences 
on both the properties and the
merger rate of binaries formed through the triple channel.
These binaries will in fact have  higher eccentricities as they enter the aLIGO band
and a higher chance of merging than what we would predict
based on the standard secular equations of motion 
 \citep{2012ApJ...757...27A,2012arXiv1211.4584K,2014ApJ...781...45A,2014MNRAS.439.1079A,2016ApJ...816...65A}.

To demonstrate the importance of the non-secular regime for the evolution of our binaries,
 we integrated all  BH triples formed in our models using the orbit-averaged equations of motion \citep{2002ApJ...578..775B}  up to a final 
 integration time of $\rm 10\ Gyr$. In total we  
 find 79 mergers, which is a factor  $2.2$ smaller than the total of $177$ mergers
we previously found.
Moreover,  the secular integrations did not produce any merging binaries with 
eccentricities larger than $\approx 0.1$ at $10\rm Hz$, 
while about $3\rm \%$ of all merging binaries that we evolved 
using the direct integrator had eccentricities larger than $(1-e_1)\lesssim 10^{-6}$ at $10\rm Hz$ (see Figures\ \ref{eccen} 
and \ref{ecc-comp}).

   \subsection{Effects due to flyby encounters} \label{flb}
In our simulations, many potential progenitors for
   BH mergers experience a collision already
during the main sequence evolution. As shown in Figure\ \ref{incl},
this reduces the number of BH triple systems with inclination $I\approx 90^{\circ}$ which can later lead to a merger.
 However, the BH triple orbits may change significantly  due to interactions
with passing stars or with the local tidal field. If such
changes are sufficiently violent to alter  the inclination after the two BHs have formed, the fraction
of triples that avoid  a phase of mass transfer throughout the stellar evolution and can merge after the three stars have become BHs  will be 
larger than the fraction  given in Table 1.

\begin{figure}
\includegraphics[angle=0,width=3.5in]{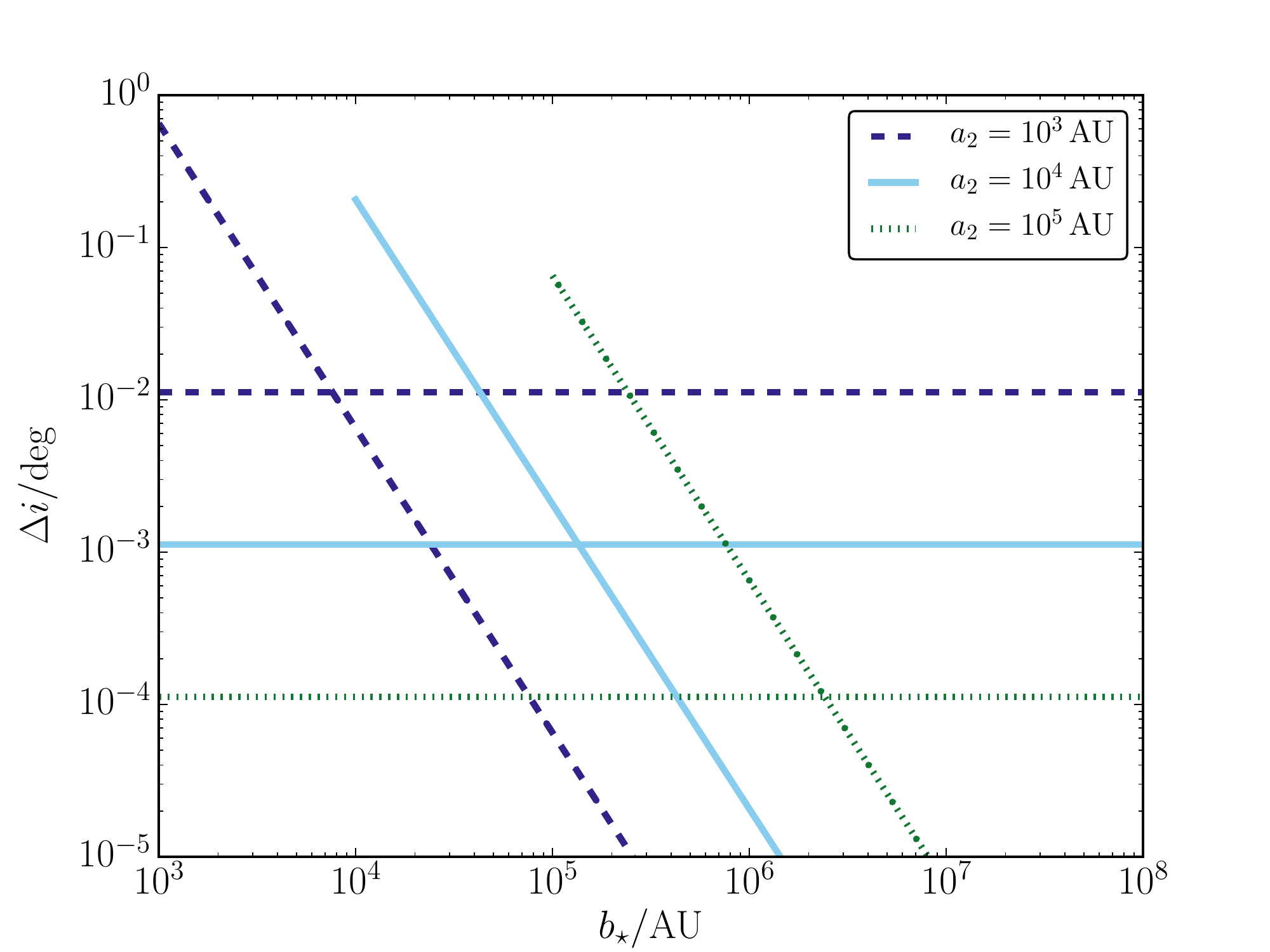}
\caption{The typical change in the inclination of the outer orbit of a typical progenitor BH triple due to a flyby with closest approach distance $b_\star$ (cf. equation~\ref{eq:Delta_i}). Three different values of the outer semi-major axis are assumed: $10^3$, $10^4$ and $10^5$ AU. The horizontal lines show $\Delta i$ averaged over $b_\star$. Refer to the
Section\ \ref{flb}  for details.}
\label{encounter_fig}
\end{figure}

\newcommand{\ve}[1]{\boldsymbol{#1}}
\newcommand{\unit}[1]{\hat{\boldsymbol{#1}}}

Here, we briefly discuss  the effect that encounters can have on the outer orbit of the BH triple, neglecting Galactic tides, which are typically less important for the systems we consider.
Generally, the effect of encounters is largest on the outer orbit of the triple. The triple BH systems which in our simulations lead to mergers typically have $a_2 \sim 10^4\,\mathrm{AU}$, with outliers at roughly $10^3$ and $10^5$ AU.  We assume the encounters are impulsive, i.e., they give a velocity kick to the outer binary. The kick $\Delta v$ to the orbital speed can be estimated as 
(e.g., \citealt[eq. 7]{2014ApJ...782...60K})
\begin{align}
\Delta v \sim \frac{3Ga_2m_\star}{b_\star^2 v_\star},
\end{align}
where $m_\star$, $v_\star$ and $b_\star$ are the mass, speed (at infinity) and closest approach distance of the perturbing star, respectively. The specific orbital angular momentum vector of the outer orbit is $\ve{j} = \ve{r} \times \ve{v}$, where $\ve{r}$ and $\ve{v}$ are the outer orbit relative orbital separation and velocity, respectively. The impulsive kick leads to an orbital velocity of $\ve{v}' = \ve{v} + \Delta \ve{v}$, whereas the position vector is assumed to be unchanged, i.e., $\ve{r}' = \ve{r}$. Assuming that the changes in $\unit{j}$, i.e., the direction of the outer orbital angular momentum, are small, we expand the expression $\unit{j}' = [\ve{r} \times (\ve{v} + \ve{\Delta v})]/|| \ve{r} \times (\ve{v} + \ve{\Delta v} )||$. Neglecting terms of order $(\Delta v)^3$ and higher, we obtain for the change of the inclination
\begin{align}
\cos(\Delta i) = \unit{j} \cdot \unit{j}' \approx 1 - \frac{1}{2} \frac{ \left (\ve{r} \times \ve{\Delta v} \right )^2}{j^2} - \frac{ \left [ \unit{j} \cdot \left ( \ve{r} \times \ve{\Delta v} \right ) \right ]^2}{j^2}.
\end{align}
Averaging the above expression over all orientations of $\ve{\Delta v}$, assuming random directions of the passing stars and keeping $\ve{r}$ and $\unit{j}$ fixed, this gives
\begin{align}
\label{eq:Delta_i}
\langle \cos(\Delta i) \rangle \approx 1 - \frac{2}{3} \frac{r^2 \Delta v^2}{j^2} \sim 1 - \frac{6G m_\star a_2^3 (1+e_2)^2}{b_\star^4 v_\star^2} \frac{m_\star}{M},
\end{align}
where $M=m_1+m_2+m_3$, $e_2$ is the outer orbital eccentricity, and we assumed that the outer binary is at apoapsis, where the effect of the encounter is strongest. 

In Fig\,\ref{encounter_fig}, we show the inclination change $\Delta i$ according to the simple estimate equation~(\ref{eq:Delta_i}) as a function of $b_\star$. We take three values of $a_2$, and set $m_1=m_2=m_3=7\,\mathrm{M}_\odot$ and $e_2=0.6$, typical for the merging systems at the moment of BH formation. The perturber parameters are set to $m_\star=0.4\,\mathrm{M}_\odot$ (a typical field star mass) and $v_\star = 40\,\mathrm{km\,s^{-1}}$ (the typical velocity dispersion in the Solar neighborhood). The lower and upper limits on $b_\star$ are set to $b_{\star,\mathrm{min}} = a_2$ and $b_{\star,\mathrm{max}} = 2\pi v_\star \sqrt{a_2^3/(GM)}$, respectively. The upper limit corresponds to the value of $b_\star$ for which the perturber passage time-scale, $\sim b_\star/v_\star$, is equal to the outer binary orbital period. If $b_\star$ is larger than $b_{\star,\mathrm{max}}$, we expect the changes to be secular (rather than impulsive); the latter are very small in low-density environments such as the field. The horizontal lines in Fig\,\ref{encounter_fig} show $\Delta i$ averaged over $b_\star$, assuming $\mathrm{d} N/\mathrm{d} b_\star \propto b_\star$. 

From Fig\,\ref{encounter_fig}, it is clear that {\it individual} encounters can only give rise to very small inclination changes, typically $\sim 10^{-3}$ degrees (for $a_2=10^4\,\mathrm{AU}$), and no larger than $\sim 1$ degree (if $a_2 = 10^5\,\mathrm{AU}$ and the encounter is deeply penetrating the outer binary). However, the {\it cumulative} change of the inclination depends on the system age, in particular, the time before the merger, which can be as long as $10^{10}\,\mathrm{yr}$. These potentially important effects will be considered in more detail in future work.

In order to demonstrate how stellar encounters  could
impact our merger rate estimates, we consider a new set of simulations where
  we take the BH triple initial conditions  from model A2  but
now sampling  the initial value of $I$ randomly  in $-1<\cos I<1$. 
The fraction of stable BH triples that lead to a merging BH binary in this new model are
$\epsilon_\mathrm{merge}=0.015$.
This corresponds to a merger rate of   
$\Gamma \approx 2.5\rm\ Gpc^{-3} yr^{-1}$.

   \subsection{BH mergers from $a_1\lesssim 10\rm AU$}
In the models of Table 1 we have set a minimum orbital separation to
avoid mass transfer based on the maximum radius of a low-mass BH
progenitor. 
However, stars more massive than $50M_\odot$   suffer from
severe wind mass losses, such that they to not reach core helium
burning before the Wolf-Rayet phase (i.e. losing their hydrogen
envelope). As a result, their maximum radius during their evolution is about an
order of magnitude smaller as for lower-mass BH progenitors. 
This means that there is a part of parameter space in which a triple
BH could form that we did not consider. 

In order to address the contribution of BH binaries that  form from massive stars with $a_1(1-e_1^2) \lesssim 2500R_\odot$, 
we evolved an additional 25k systems in which 
we take $a_{2,\rm max}=5\times10^6R_\odot$, $m_{3,\rm min}=0.1\ M_\odot$, assume no-kicks as in model A1 but now
 setting $250\lesssim a_1(1-e_1^2)\lesssim 2500R_\odot$, and  $(m_1,m_2)>50 M_\odot$.  
For this set of parameters  $\epsilon_{\rm p-space}\approx 0.3$, similar to our standard models.
These simulations resulted in 
$\epsilon_\mathrm{3{BH}}=0.11$ and $\epsilon_\mathrm{merge}=0.01$, giving 
roughly  the same fraction of merging binaries (i.e., $\epsilon_\mathrm{3{BH}} \epsilon_\mathrm{merge}$) as model A1.
 However, the fraction of all triples with $(m_1,m_2)>50 M_\odot$ is about 10 times less
 than triples with $(m_1,m_2)>22 M_\odot$, 
suggesting a total merger rate of $\Gamma\lesssim 0.1\rm Gpc^{-3}yr^{-1}$ and 
 a contribution at $10\%$ level in our models.

 \section{Conclusions}\label{concl}
 A large fraction ($\approx 25\%$) of massive stars are observed to be in triple systems \citep{2014ApJS..215...15S}.
In this paper we have studied how the evolution of  massive  triples can lead to the formation of 
BH  triple systems and how their subsequent dynamical evolution can result
into the merger of  two BHs.
We study this problem using a combination of high-precision direct integrations \citep{2008AJ....135.2398M} and  a code
that combines secular three-body dynamics with stellar evolution and their mutual influences \citep{2016ComAC...3....6T}.
Our approach allows us to make reliable predictions about the properties of the newly formed BH triples,
and  about the rates and properties of the merging binaries.
The main conclusions of our study are summarized below:\\
\begin{itemize}
\item[1)] In our models with no natal kicks, about
$ 20 \%$ of the massive triple stars evolve into
 a hierarchical  BH triple system;  in models that  include non-zero natal kicks this percentage is smaller, $\approx 5\rm \%$, 
as many triples are disrupted during BH formation.
About 
 $\approx 1\rm \%$ of the BH triples that are formed  in our models
  produce a merging BH binary. This latter fraction is roughly independent on the assumed distribution of 
natal kick velocities.
     \\
 \item[2)] 
 The majority (132 out of 177) of BH mergers in our models are formed 
 through a complex non-secular dynamical  evolution which cannot be adequately modeled 
 using the standard (double-averaged) secular perturbation theory, even at the octupole level of approximation.
 Compared to direct 
 three-body integrations, secular calculations underpredict  the  number of BH mergers by a factor $\approx 2$ as well as the
 eccentricity of the merging binaries.
 \\
\item[3)] 
We estimate the rate of BH mergers in isolated 
triples in the field to be in the range  $(0.3- 1.3)\ \rm Gpc^{-3}yr^{-1}$.
If the orbital inclinations of the BHs are efficiently randomized, for example as 
a consequence of interactions with passing field stars,
 the resulting merger rate could 
be as large as $\approx 2.5\ \rm Gpc^{-3}yr^{-1}$.
  \item[4)] 
Inspiraling  BH binaries formed in field triples have significantly higher eccentricities
than those formed through the evolution of field binaries or via dynamical interactions in dense star  clusters.
A few per cent of merging binaries in our models enter the aLIGO frequency band with $(1-e_1)\lesssim 10^{-6}$.
We conclude that measured eccentricities could  provide a way to uniquely identify  binary mergers formed through the evolution of massive triple stars. 
\end{itemize}

\bigskip
 FA acknowledges  support  from  a  CIERA  postdoctoral  fellowship
at  Northwestern  University. ST acknowledges support from the
NetherlandsResearch Council NWO (grant VENI [nr. 639.041.645]).
ASH acknowledges support from the Institute for Advanced Study, and NASA NNX14AM24G grant.

 \enddocument